\documentclass[fleqn,usenatbib]{mnras}

\usepackage{graphicx}	
\usepackage{amsmath}	
\usepackage{amssymb}	
\usepackage{longtable}	
\usepackage{subfig}

\usepackage[T1]{fontenc}
\usepackage{ae,aecompl}
\usepackage{hyperref}

\usepackage{graphicx}	
\usepackage{amsmath}	
\usepackage{amssymb}	

\title[KiDS-AMICO galaxy clusters]{AMICO galaxy clusters in KiDS-DR3: galaxy population properties and their redshift dependence}
 
\author[M. Radovich et al.]{Mario Radovich$^{1}$\thanks{E-mail: mario.radovich@inaf.it},
	Crescenzo Tortora$^{2,3}$,
 	 Fabio Bellagamba$^{4,5}$,
 	Matteo Maturi$^{6,7}$,
\newauthor 	
 	Lauro Moscardini$^{4,5,8}$,
	 Emanuella Puddu$^{2}$,
	Mauro Roncarelli$^{4,5}$,
 	 \newauthor
	Nivya Roy$^{9}$,
 	 Sandro Bardelli$^{5}$, 
	Federico Marulli$^{4,5,8}$,
	Mauro Sereno$^{4,8}$,
	\newauthor
	Fedor Getman$^{2}$,
	Nicola R. Napolitano$^{9,2}$
 \\ \\
	$^{1}$ INAF - Osservatorio Astronomico di Padova, vicolo dell'Osservatorio 5, I-35122 Padova, Italy\\
	$^2$  INAF - Osservatorio Astronomico di Capodimonte, Salita Moiariello 16, 80131 - Napoli, Italy\\
    $^3$ INAF - Osservatorio Astrofisico di Arcetri, Largo Enrico Fermi 5, I-50125 Firenze, Italy\\
	$^{4}$  Dipartimento di Fisica e Astronomia - Alma Mater Studiorum Universit\`{a} di Bologna, via Piero Gobetti 93/2,\\ I-40129 Bologna, Italy\\
	$^{5}$ INAF - Osservatorio di Astrofisica e Scienza dello Spazio di Bologna, via Piero Gobetti 93/3, I-40129 Bologna, Italy\\
	$^{6}$ Zentrum f\"ur Astronomie, Universitat\"at Heidelberg, Philosophenweg 12, D-69120 Heidelberg, Germany\\
    $^{7}$ ITP, Universität Heidelberg, Philosophenweg 16, D-69120 Heidelberg, Germany\\
	$^{8}$ INFN - Sezione di Bologna, viale Berti-Pichat 6/2, I-40127 Bologna, Italy \\
	$^9$ School of Physics and Astronomy, Sun Yat-sen University Zhuhai Campus, Daxue Road 2, 519082 \\ 
	     Tangjia, Zhuhai, Guangdong, P.R. China
}

\date{Accepted XXX. Received YYY; in original form ZZZ}

\pubyear{2020}

\begin{document}

\newcommand{\bcgmgdr}[1]{1400\ }
\def\lephare{\mbox{\textsc{le phare}}}
\def\AMICO{\mbox{\textsc{AMICO}}}
\def\ezgal{\mbox{\textsc{ezgal}}}
\def\bpz{\mbox{\textsc{bpz}}}
\def\galcol{\mbox{\textsc{col}}}
 
\label{firstpage}
\pagerange{\pageref{firstpage}--\pageref{lastpage}}
\maketitle

\begin{abstract}
A catalogue of galaxy clusters was obtained in an area  of 414 $\deg^2$ up to a redshift $z\sim0.8$ from the Data Release 3 of the Kilo-Degree Survey (KiDS-DR3), using  the  Adaptive Matched Identifier of Clustered Objects (\AMICO) algorithm. The catalogue and  the calibration of the richness-mass relation were presented in two companion papers. Here we  describe the selection of the cluster central galaxy and  the classification of blue and red cluster members, and analyze the main cluster properties, such as the red/blue fraction, cluster mass, brightness and stellar mass of the central galaxy, and their dependence on redshift and cluster richness. We use the Illustris-TNG simulation, which represents the state-of-the-art cosmological  simulation of galaxy formation, as a benchmark for the interpretation of the results.
 A good  agreement with simulations is found at low redshifts ($z \le 0.4$), while at higher redshifts the simulations indicate a lower fraction of blue galaxies than what found in the KiDS-\AMICO\ catalogue: we argue that this may be due to an underestimate of star-forming galaxies in the simulations. The selection of clusters with a larger magnitude difference between the two brightest central galaxies, which may indicate a more relaxed cluster dynamical status, improves the agreement between the observed and simulated cluster mass and stellar mass of the central galaxy. We also find that at a given cluster mass the stellar mass of blue central galaxies is lower than that of the red ones.
\end{abstract}




\section{Introduction}

Being the most massive collapsed structures in the Universe, 
galaxy clusters provide a fundamental tool to study the effect of massive dark matter halos on the  properties of their member galaxies \citep{2018ARA&A..56..435W} and how they evolve with redshift. A remarkable progress in our understanding of the cluster formation and evolution was achieved in the last decades \citep[see e.g.][]{1984ARA&A..22..185D, 2011ASL.....4..204B, 2012ARA&A..50..353K}, but it is clear that clusters provide  a complex environment where a  variety of physical processes take place, such as star formation, AGN (Active Galactic Nucleus) feedback, tidal stripping: this is particularly true for the galaxy located at the centre of the cluster halo \citep{2007MNRAS.375....2D,2014MNRAS.443.1500M}, which  is often, but not always \citep{2015MNRAS.452..998H, 2018MNRAS.480.2689H},  the brightest cluster galaxy (BCG).
Cluster formation hierarchical models \citep[see e.g.][]{2007MNRAS.375....2D,2010ApJ...710..903M,2012ARA&A..50..353K,2018AstL...44....8K,2020A&A...634A.135G} predict  a strong connection between the cluster halo and its BCG. In this scenario, the stellar mass of the BCG is closely related to the mass of the dark matter halo in which it formed, with the most massive halos  hosting the most massive BCGs: as the BCG continues to grow through merging with the surrounding satellite galaxies, its size, luminosity and  stellar mass, as well as the magnitude difference with respect to other nearby cluster members, increase
\citep[see e.g.][]{2007MNRAS.379..867V,2007AJ....133.1741B,2019A&A...631A.175E}.
On the other hand, it is now becoming clear that large-scale environment also plays an important role: clusters with a disturbed dynamical status due to major mergers with smaller clusters have revealed to be more frequent than expected from previous observations \citep{2013MNRAS.436..275W}. Mergers are predicted to increase the cluster mass  \citep{2018MNRAS.478.5473L} and possibly  modify the relation between the cluster halo and BCG mass  \citep{2016MNRAS.462.4141L,2018MNRAS.478.5473L} observed in relaxed clusters, and affect the BCG luminosity in particular at high redshifts  \citep{2020MNRAS.495..705Z}.

Recent cosmological hydrodynamic simulations  such as those provided by the Illustris Project   \citep{2014Natur.509..177V,2015A&C....13...12N} and its latest release, Illustris-TNG \citep{2018MNRAS.480.5113M,2018MNRAS.477.1206N,2018MNRAS.475..676S,2018MNRAS.475..648P,2018MNRAS.475..624N,2019ComAC...6....2N}  are now enabling  the possibility to match reasonably well the cluster observables, thanks to the  combination  of large volume and high particle resolution, even if they may not be yet able to describe all the complex physical processes that shape clusters \citep[see e.g.][]{2018MNRAS.481.1809B}.

On the side of observations, wide-field surveys  such as the Sloan Digital Sky Survey \citep[SDSS, ][]{2000AJ....120.1579Y}, the Kilo-Degree Survey \citep[KiDS, ][]{2013Msngr.154...44D}, the Dark Energy Survey \citep[DES, ][]{des16}, the Hyper Suprime-Cam Survey \citep[HSC, ][]{2018PASJ...70S...4A}, provide  multi-wavelength photometry of galaxies in individual clusters: the selection of a well controlled but statistically significant number of clusters in  a wide range in mass and redshift is required to study in detail how this complex interplay of physical phenomena takes place. This enables us to understand if and how different observables (e.g. the cluster halo mass, the brightness of the central galaxy, star formation in galaxy clusters vs. the distance from the cluster centre, etc.) are related, and how they evolve in cosmic time.

Several algorithms to efficiently search for galaxy clusters have been developed  \citep[see e.g.][and references therein for a recent review]{2019A&A...627A..23E}, based either on the fact that early-type galaxies  occupy a well defined  position (the red sequence) in the colour-magnitude space at the cluster redshift \citep[e.g. {\em redMaPPer} and CAMIRA:][respectively]{2014ApJ...785..104R, 2018PASJ...70S..20O}, or on the detection with optimal matched filters \citep[see e.g.][]{2011MNRAS.413.1145B} of the galaxy overdensities   which are the signatures of galaxy clusters. These produced different catalogues of clusters, allowing to study the properties of their member galaxies  \citep[see e.g.][]{2018MNRAS.481.4158W,2018A&A...613A..67S,2018PASJ...70S..24N, 2020ApJ...897...15T}, though a comparison of the results is challenging due to the different survey properties (area and depth) and algorithm selections.

The Adaptive Matched Identifier of Clustered Objects (\AMICO) algorithm  \citep{2018MNRAS.473.5221B} was used to search for galaxy clusters in the Kilo-Degree Survey \citep{2017A&A...598A.107R, 2019MNRAS.485..498M}. This algorithm 
does not make an explicit use of colours, in contrast to algorithms based on the detection of a red sequence. This is of great benefit when studying the galaxy population of clusters because the sample selection is not so determined by a specific set of galaxies,
allowing for instance to detect clusters with bluer populations. In this paper 
we analyze the fraction of red and blue galaxies in the KiDS-\AMICO\  clusters and its dependence on  cluster mass and redshift, and the properties of red and blue BCGs. In a separate paper (Puddu et al., in preparation), we address the dependence of  the red and blue cluster galaxies luminosity function on redshift and mass.

The paper is structured as follows. A short summary of the KiDS dataset and of the KiDS-\AMICO\ cluster catalogue is given in Section~\ref{sec:data} and Section~\ref{sec:amico} respectively. Section~\ref{sec:amico} also provides a description of the method adopted for the selection of the BCG, and of the validation, based on available spectroscopic redshifts,  of the \AMICO\ membership probabilities.  Section~\ref{sec:red_blue} describes the selection  of the red and blue cluster members and their fractions as a function of  redshift and distance from the cluster centre are discussed, and compares the KiDS-\AMICO\ and {\em redMaPPer} cluster detections  in the same areas and within the same cluster mass and redshift range. The  properties of the BCGs, such as their luminosity and  stellar mass  are described in Section~\ref{sec:other}. Section~\ref{sec:comp2TNG} compares the results obtained in this paper with the Illustris-TNG300-1 simulations. Results are summarized and discussed in Section~\ref{sec:conclusions}.

The cosmology concordance model was adopted throughout the paper: $H_0$ = 70 km s$^{-1}$ Mpc$^{-1}$, $\Omega_m$ = 0.3, $\Omega_\Lambda$ = 0.7.

\section{The Kilo-Degree Survey}
\label{sec:data}

The Kilo-Degree Survey \citep[KiDS,][]{2013Msngr.154...44D} is an ESO Public Survey observing  with the OmegaCAM camera on the ESO VLT Survey telescope (VST) in the $ugri$ bands an area of 1350 deg$^2$ distributed in two stripes, one equatorial (KiDS-N) and the other towards the South Galactic Pole (KiDS-S). Three main Data Releases are currently available: KiDS-DR2, covering $\sim$100 deg$^2$ \citep{2015A&A...582A..62D}; KiDS-DR3, extending to 440 deg$^2$ \citep{2017A&A...604A.134D}; KiDS-DR4 \citep{2019A&A...625A...2K}, reaching 1000 deg$^2$.
The supplementary catalogue KV450  \citep{KV450a} joined for the first time photometry from KiDS and near-infrared photometry ($ZYJHK_s$) from the parallel ESO Public Survey VISTA Kilo-Degree Infrared Survey (VIKING), for the KiDS-DR3 area. Finally, spectroscopic redshifts for the brightest ($r<20$ mag) galaxies in KiDS are available through the overlap with the Sloan Digital Sky Survey \citep[SDSS, ][]{2000AJ....120.1579Y} and the Galaxy and Mass Assembly survey  \citep[GAMA, ][]{2009A&G....50e..12D} in KIDS-N.

\citet{2018MNRAS.480.1057R} derived the structural parameters (S\'ersic index and effective radius) of the brightest (S/N>50)\footnote{Here the signal-to-noise was defined as the inverse error of $r-$band Kron-like magnitudes in SExtractor \citep{Bertin_Arnouts96_SEx}: S/N=1.086/MAGERR\_AUTO\_r.}  galaxies in  KiDS-DR2: the S\'ersic index and effective size were computed fitting  their $gri$ images with S\'ersic models.
The same analysis was extended to KiDS-DR3 and will be presented in a separate paper. In order to select the best fit structural parameters, here we  imposed the constraint $\chi^2 < 1.3$. As discussed by \citet{2018MNRAS.480.1057R}, larger values of 
$\chi^{2}$ correspond to strong residuals, often associated to spiral arms.
Moreover, as described by \citet{2018MNRAS.481.4728T}
stellar masses were computed  from $ugri$ photometry in KiDS-DR3 using the code \lephare\ 
\citep{Arnouts+99, Ilbert+06}, which fits the KiDS photometry to a stellar population synthesis (SPS) theoretical model. Single burst models
from \citet{2003MNRAS.344.1000B} were used, covering all the range of
available metallicities ($0.005 \leq Z/Z_\odot \leq 2.5$), with $\rm
age$ smaller than the age of the Universe at the redshift of the galaxy (with a maximum value at $z=0$ of $13\, \rm
Gyr$) and a \cite{Chabrier01} IMF. Age and metallicity were left free to vary in the fitting
procedure. Models were redshifted using the photometric
redshifts derived with the same code, \bpz\ \citep{2000ApJ...536..571B}, used to compute photometric redshifts in KiDS. $ugri$ magnitudes were measured within a circular aperture of diameter 6 arcsec (and related $1\, \sigma$
uncertainties $\delta u$, $\delta g$, $\delta r$ and $\delta i$) and corrected for Galactic extinction using the map in
\cite{Schlafly_Finkbeiner11}. The $r-$band Kron-like magnitudes (MAG\_AUTO\_r) were used to correct the stellar mass outcomes
of \lephare\ for missing flux. Calibration zero-point errors, 
$(\Delta u_{zp}, \, \Delta g_{zp}, \, \Delta r_{zp}, \, \Delta
i_{zp}) = (0.075, \, 0.074, \, 0.029, \, 0.055)$, were added in quadrature to the
uncertainties of the magnitudes derived from SExtractor.

\section{The KiDS-AMICO DR3 cluster catalogue }
\label{sec:amico}

Differently from other cluster search codes as e.g. {\em redMaPPer} \citep{2014ApJ...785..104R} and \texttt{CAMIRA}
\citep{2018PASJ...70S..20O}, the 
\AMICO\ algorithm is not based on the detection of the cluster red-sequence: instead, an optimal matched filter is applied 
to a catalogue with coordinates, photometric redshifts and magnitudes, allowing to select galaxy overdensities tracing the presence of galaxy clusters \citep[for details on the algorithm see][and references therein]{2019MNRAS.485..498M}. The code produces a list of clusters with their centres ($\mathbf{x}$), redshift ($z_{\rm cl}$), significance (SN) and the amplitude ($A$).  For each cluster candidate, a catalogue with the membership probabilities ($p_{\rm memb}$) of galaxies is also available. A first catalogue of galaxy clusters detected in KiDS-DR2 was presented in \citet{2017A&A...598A.107R}. Later, the wider area available in KiDS-DR3 and the new features implemented in the detection algorithm enabled a new analysis, producing a catalogue  of 7988 clusters in the redshift range 0.1$< z < 0.8$ over an area of 414 $\deg^2$ after removing areas around bright saturated stars. As described in \citet{2019MNRAS.485..498M},  a cut SN$>3.5$ was adopted to 
minimize the number of spurious detections.

\subsection{Richness and mass }

 A full analysis of the properties of this catalogue was presented in  \citet{2019MNRAS.485..498M}, including the  characterization of uncertainties on the \AMICO\ cluster parameters, the evaluation of purity and completeness, and   the selection function. Moreover, two new richness parameters ($\lambda$ and $\lambda^*$), derived from the \AMICO\ membership probabilities, were introduced. In particular, the {\em intrinsic richness} $\lambda^*$  was designed to reduce its dependence on the redshift, compared to the {\em apparent richness} $\lambda$.

 Based on the shear measurements in KiDS-DR3 \citep{2015MNRAS.454.3500K,2017MNRAS.465.1454H},
 \citet{2019MNRAS.484.1598B} made a weak lensing stacked analysis to
 calibrate the relation between $\lambda_*$  and the cluster mass $M_{200}$, defined as the mass within the radius $R_{200}$ where the mean density is 200 the critical density of the Universe at that redshift \citep[Eq.~31 in ][]{2019MNRAS.484.1598B}: 
\begin{equation}
\log{\frac{M_{200}}{10^{14} M_\odot h^{-1} }} = \alpha + \beta \log {\frac{\lambda^*}{\lambda^*_{\rm piv}} +\gamma \log{\frac{E(z)}{E(z_{\rm piv)}}} }, \label{eq:mscale}
\end{equation}
where $E(z)=H(z)/H_0$, $\alpha=0.004\pm0.038$, $\beta=1.71\pm0.08$,  $\gamma=-1.33\pm0.64$,
$\lambda^*_{\rm piv}=30$ and $z_{\rm piv}=0.35$.

\begin{figure} 
	\includegraphics[clip,width=1.0\linewidth]{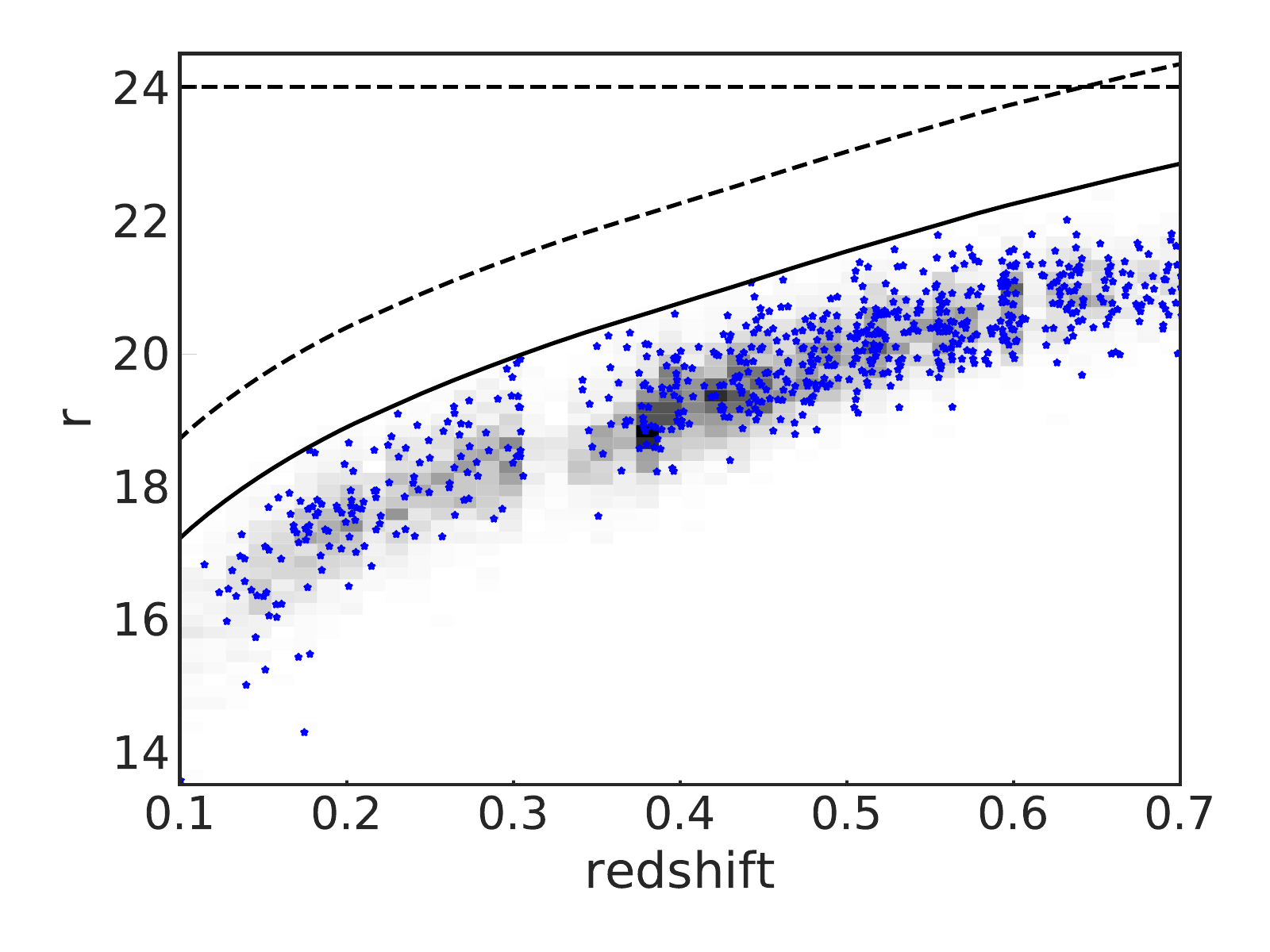}
	\caption{$r-$band apparent magnitudes of the BCGs (blue dots display late-type BCGs). The solid curve shows the $m_*$ values adopted as a lower brightness limit for the BCGs. The dashed curve ($m_*+1.5$)  is the lower limit adopted for the selection of member galaxies in the present analaysis; the dashed horizontal line is the cut ($r$=24 mag) adopted to select galaxies in \AMICO. 
	The gap at $z\sim0.35$ corresponds to the redshift interval with a reduced coverage of features in the spectral energy distribution by the  $ugri$ filters, as discussed in \citet{2019MNRAS.485..498M}.
	\label{fig:bcg_sel}}
\end{figure}

\begin{figure} 
	\includegraphics[clip,width=1.0\linewidth]{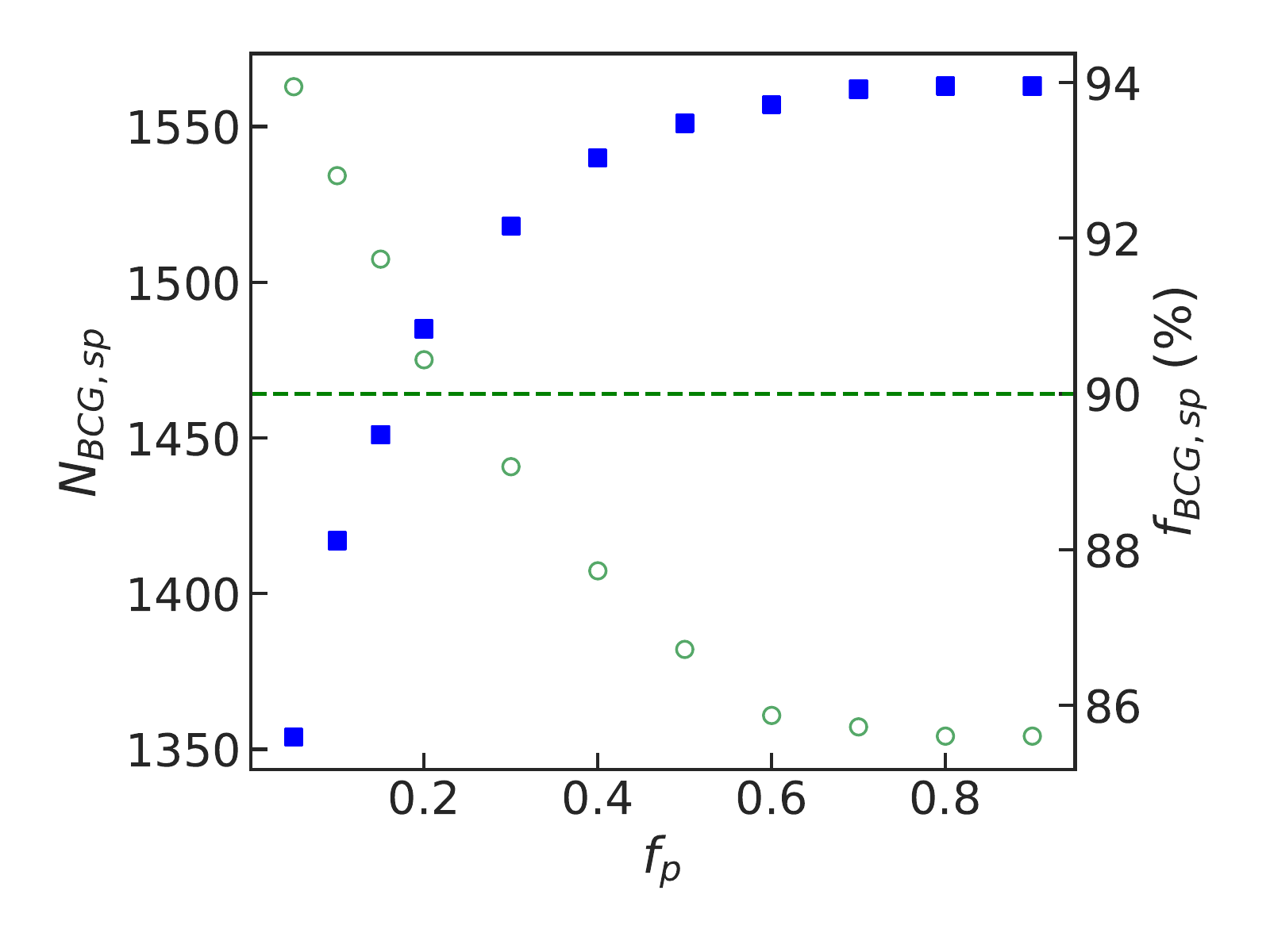}
	\caption{The plot displays, for different values of $f_p$: {\em filled squares - }the number of BCGs
		with spectroscopic redshifts, left axis); {\em empty circles - }the fraction of BCGs with cluster redshifts in agreement with spectroscopic redshifts as defined in the text ($f_{\rm BCG,sp}$, right axis). The dashed line is the threshold fraction adopted to select the optimal values of $f_p$. \label{fig:bcg_train}}
\end{figure}

\begin{figure}
	\includegraphics[clip,width=.9\linewidth]{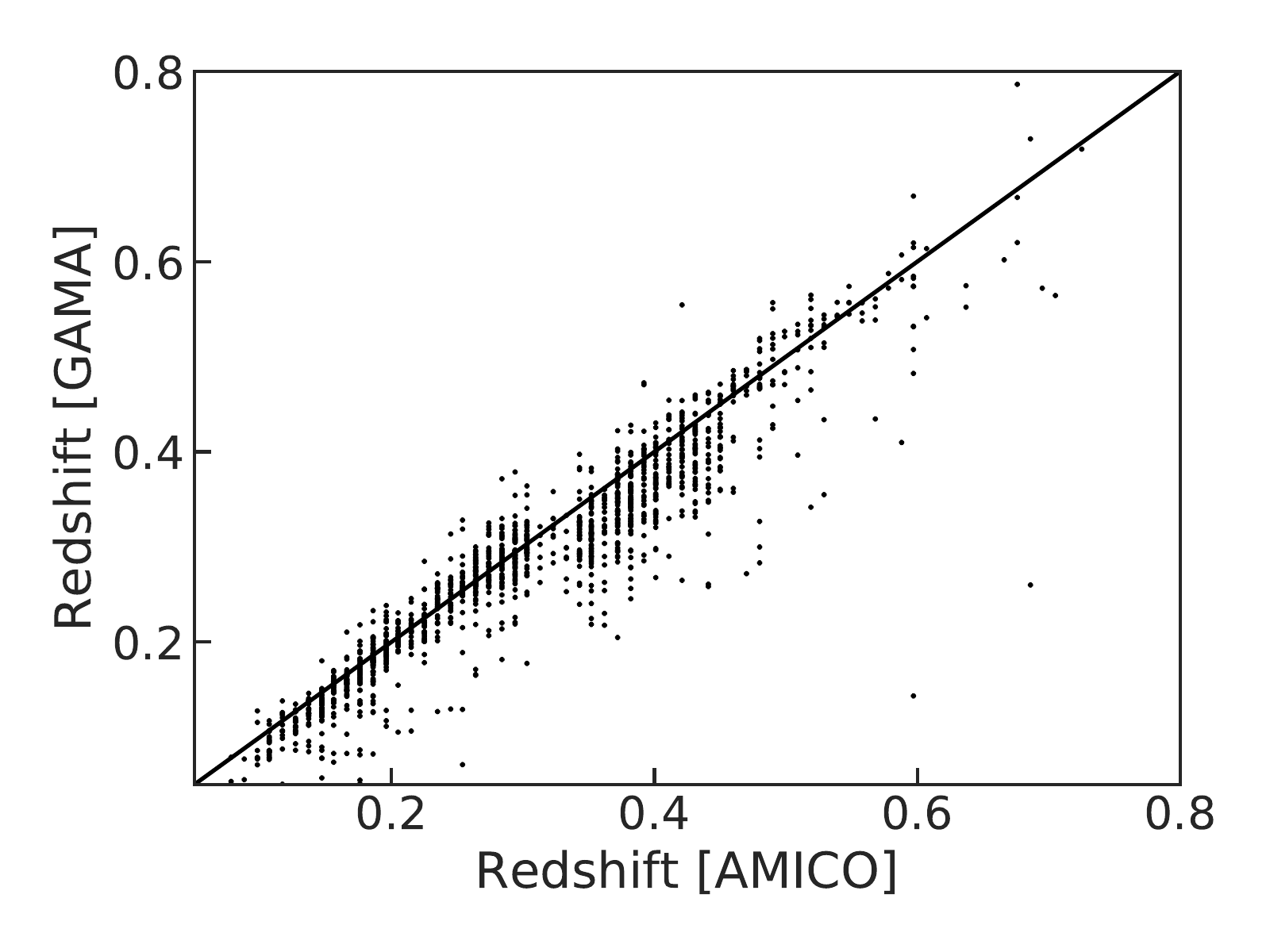}
	\caption{Comparison of \AMICO\ and spectroscopic (GAMA/SDSS) redshifts for the galaxies identified as the BCGs. The solid line displays the 1-to-1 relation as a reference. \label{fig:redshifts}}
\end{figure}

\begin{figure} 
	\includegraphics[clip,width=.9\linewidth]{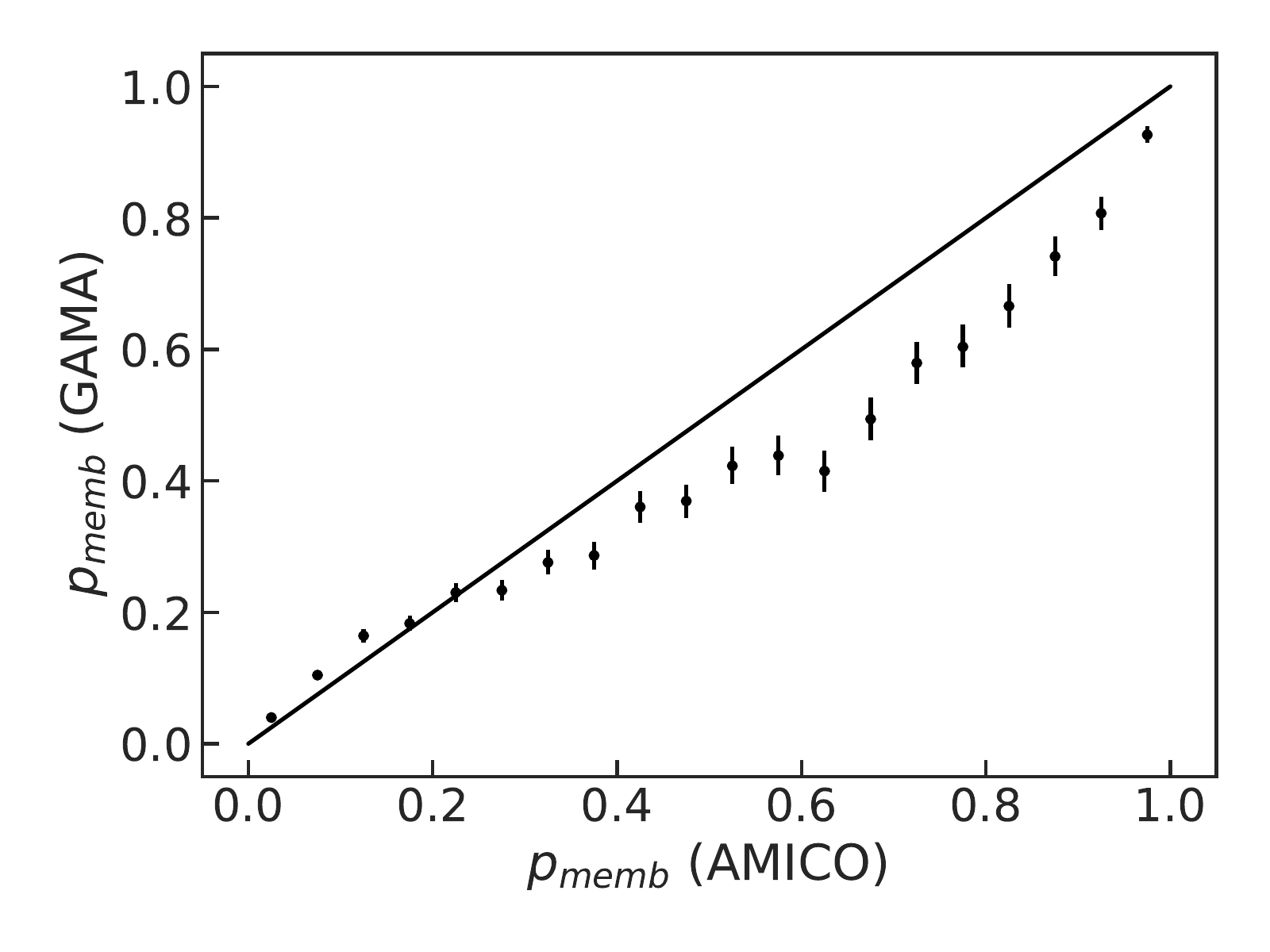}
	\caption{\AMICO\ membership probabilities compared to the fraction of members computed from the galaxies with spectroscopic redshifts. The solid line displays the 1-to-1 relation as a reference.\label{fig:zmemprob}}
\end{figure}

\subsection{Selection of the BGC}
\label{sec:bcg_selection}

Identifying the central galaxies in clusters is fundamental as their position at the centre of the dark matter halos from which the cluster formed leads to peculiar properties compared to the other cluster galaxies.
However, as discussed by e.g. \citet{2015MNRAS.452..998H} and \citet{2018MNRAS.480.2689H},  the cluster central galaxy does not necessarily coincide with the brightest galaxy in the cluster. For this reason,   we consider not only the galaxy luminosity, but also its distance from the initial \AMICO\ centre and the membership probability to select the BCG. This  is done as follows.  For each cluster we first compute
the  characteristic magnitude $m_*$ (see Figure~\ref{fig:bcg_sel}), defined as the absolute $r$--band magnitude $M_r^*(z=0.1) - 5 \log{h}$, where $M_r^*(z=0.1)=-20.44$ \citep{2003ApJ...592..819B}, transformed to the observed apparent magnitude at the cluster redshift. This is done with the \ezgal\ code \citep{2012ascl.soft08021M}, taking as input the \citet{2003MNRAS.344.1000B}  population synthesis model with metallicities of $Z/Z_\odot \sim 1$ \citep{2012PASP..124..606M}. The model is calibrated normalizing the model magnitudes to the median observed values, at a redshift $z\sim 0.15$. 

The procedure first extracts the galaxies with an $r$-band magnitude brighter than $m_*$, and  at the same time with $|p_{\rm memb} - p^{\rm max}_{\rm memb}|/p^{\rm max}_{\rm memb} < f_p$ to remove bright, foreground galaxies not belonging to the cluster; $f_p$ is a constant  whose value is assigned as described below, 
$p^{\rm max}_{\rm memb}$ is the highest value of the membership probability in each cluster.
We then further select those galaxies whose distance from the  \AMICO\  cluster centre ($\mathbf{x}$) is within its uncertainty in angular position, $\Delta \mathbf{x}$: as  described in \citet{2019MNRAS.485..498M}, $\Delta \mathbf{x}$ is a function of the cluster redshift, decreasing from $\sim$ 3 arcmin at $z<0.1$ to $\sim$ 0.35 arcmin at $z>0.45$. These galaxies are then  sorted  in decreasing membership probability. The BCG is defined as the brightest galaxy in the $r$-band among the first 5 sorted galaxies; if the difference in magnitude between the first two brightest galaxies is lower than $\pm$ 0.1 mag, the galaxy with the highest membership probability is selected. If there is no galaxy within $\Delta \mathbf{x}$, we extend the search to $2 \times \Delta \mathbf{x}$, and so on up to a maximum distance of $5 \times \Delta \mathbf{x}$. The second brightest galaxies selected in the same way is also stored, to compute the cluster magnitude gap.

To find the optimal choice for $f_p$, we proceed as follows.  The value of $f_p$ is varied between 0.05 and 0.9, and each time the BCGs are identified as descrived above.  Their positions are  matched to the GAMA-DR3 catalogue \citep{2018MNRAS.474.3875B}, producing $N_{\rm BCG, sp}$ galaxies with GAMA or SDSS spectroscopic redshifts. 
We then discard those BCGs with a spectroscopic redshift, $z_{\rm sp}$, non compatible with the one of the cluster detection, $z_{\rm cl}$, i.e with $|z_{\rm cl}-z_{\rm sp}|/(1+z_{\rm sp}) > 0.1$, obtaining $N^*_{\rm BCG,sp}$ galaxies. The fraction is  $f_{\rm BCG,sp}=N^*_{\rm BCG,sp}/N_{\rm BCG, sp}$.  
The result is displayed in Figure~\ref{fig:bcg_train}, showing that  an 
increase in $f_p$ produces  more  BCGs with spectroscopic redshifts (higher $N_{\rm BCG, sp}$), but also more mismatches between spectroscopic and cluster redshifts (lower $f_{\rm BCG,sp}$). We defined  $f_{\rm BCG,sp}=0.9$ as a threshold, which is reached when $f_p = 0.2$. This is  the value for $f_p$ adopted in the following analysis.

\subsection{Validation of the membership probability }
\label{sec:membership}

Spectroscopic redshifts are also used to validate the \AMICO\  membership probability. Since for each cluster  few ($<$ 10) members with spectroscopic redshifts are expected to be found, to this end we make a stacked analysis \citep[see e.g.][]{2015MNRAS.453...38R} comparing the average spectroscopic and \AMICO\ membership probabilities for the sample of clusters with spectroscopic redshifts. The spectroscopic membership rate is derived with the following approach:
\begin {enumerate}
\item 
We select the $\sim$ 1400 clusters for which 
it is possible to assign a spectroscopic redshift to the galaxy identified as the BCG: this value is adopted as the {\em initial} cluster redshift ($z_{sp,0}$).  Figure~\ref{fig:redshifts} compares this redshift with the \AMICO\ cluster redshift. 
\item For each cluster, we select all galaxies with spectroscopic redshifts ($z_{sp,i}$), within a distance of 0.5 $h^{-1}$ Mpc and 5 arcmin from the \AMICO\ cluster centre. {\em Spectroscopic members} are defined as  galaxies with a velocity offset $|\frac{z_{sp,i}-z_{sp,0}}{1+z_{sp,0}}| < f \sigma_v$, where $f$ is a threshold ($f=3$) and  $\sigma_v$ = 1500 km\ s$^{-1}$ for the first iteration. 
\item New values of $z_{sp,0}$ and $\sigma_v$ are computed as the biweight average and standard deviation of the redshifts for these galaxies, and the procedure is repeated until convergence. We verify that the results do not change significantly with different choices of $f$ and the initial velocity dispersion. We finally select $\sim$ 760 clusters with at least 3 spectroscopic members. For these clusters, of the  $\sim$50\,000 galaxies for which both a membership probability and a spectroscopic redshift are assigned, $\sim$ 12\,000 are classified as {\em spectroscopic members} based on the above criteria (the average number of spectroscopic members per cluster is 16).
\item  The \AMICO\ membership probabilities are binned in steps of 0.05: in each bin we define a spectroscopic membership rate $n_{sp} = N_{mem}/N_{tot}$, and compare it with the \AMICO\ membership probability.
Error bars on spectroscopic memberships are obtained by bootstrapping: data are randomly resampled  with substitution 10\,000 times, spectroscopic memberships are computed  and the 5\% and 95\% percentiles are used as lower or upper limit.\footnote{The same approach is adopted to obtain confidence intervals throughout the paper.}
\end{enumerate} 

 The results,  displayed in Figure~\ref{fig:zmemprob}, are in good agreement with the comparison with simulated mock catalogues discussed in \citet[][Figure~18]{2018MNRAS.473.5221B}, where deviations between expected and \AMICO\ membership probabilities are explained by mismatches between the true and model size of the clusters and to miscenterings of the halo positions.

\begin{figure}
	\includegraphics[clip,width=1.05\linewidth]{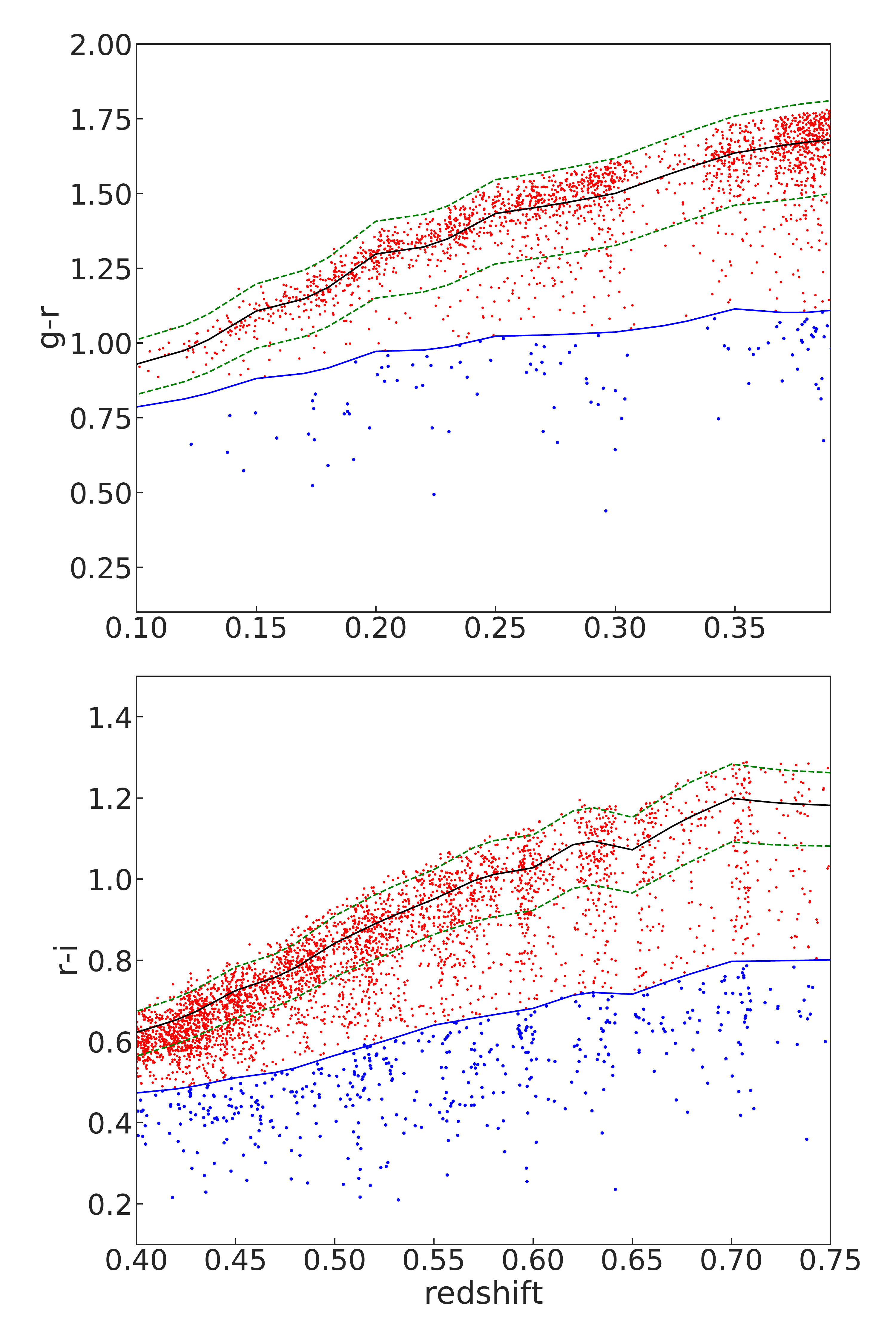}
	\caption{Observed colours of the BCGs are compared to colours for the {\em E} (red line) and {\em Sa} (blue line) models defined in the text; the green dashed lines show the elliptical models at metallicities $Z/Z_\odot$ = [0.5, 2].
		'Red' and 'blue' BCGs are displayed as red and blue dots respectively.
		\label{fig:colBCG}}
\end{figure}

\begin{figure}
	\includegraphics[clip,width=1\linewidth]{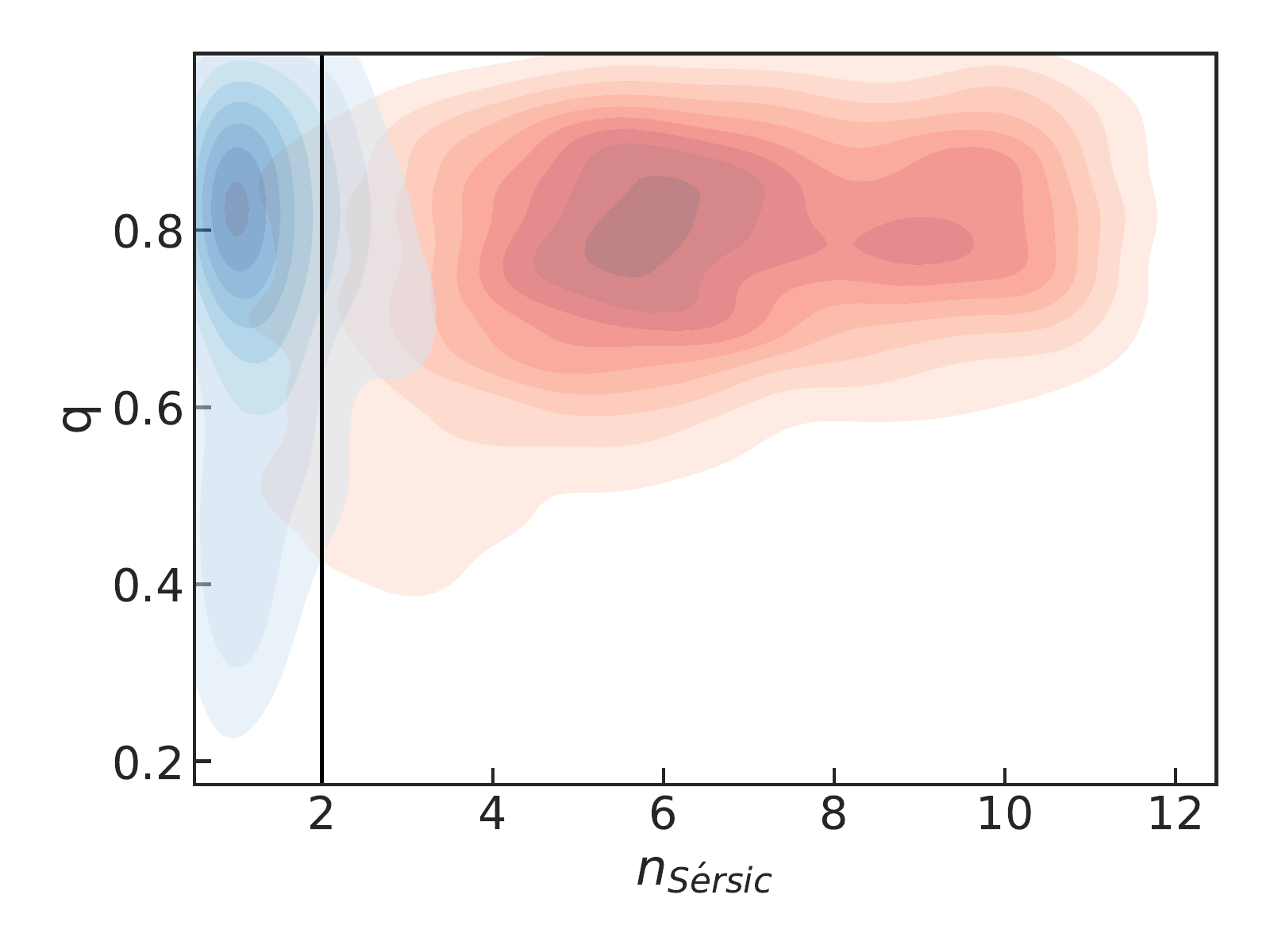}
	\includegraphics[clip,width=1\linewidth]{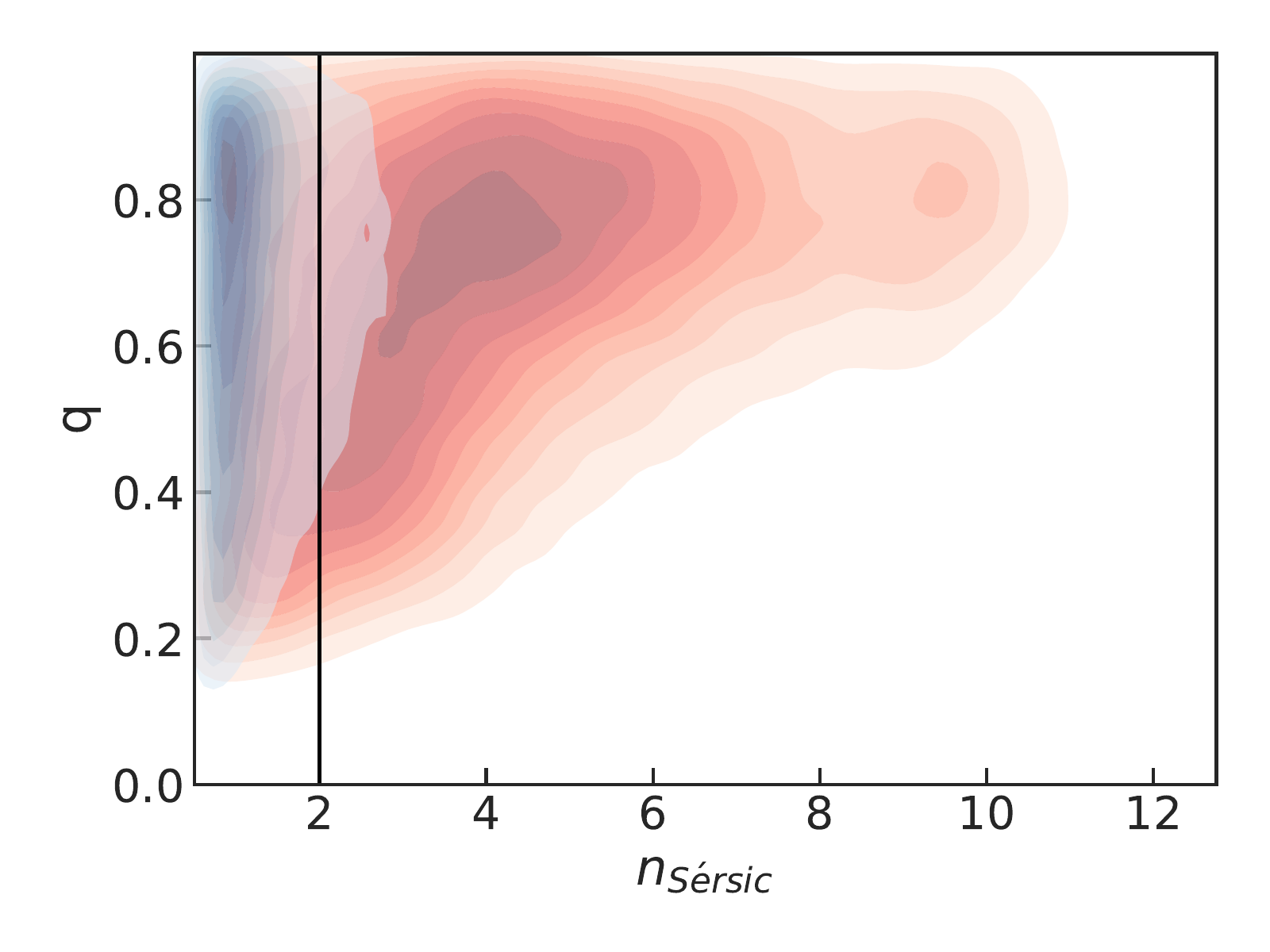}
	\caption{Distribution of the axial ratio ($q$) and of  the S\'ersic index for the galaxies classified as blue and red based on their colours. The plot above shows BCGs only, in the plot below all galaxies brighter than $r=20$ mag are displayed. The vertical line ($n_{S\acute{e}rsic}$=2) is the limit adopted to separate red and blue galaxies based on structural parameters.\label{fig:bcg_sersic}}
\end{figure}

\begin{figure*}
	\includegraphics[clip,viewport=25bp 0bp 864bp 288bp, width=1.1\linewidth]{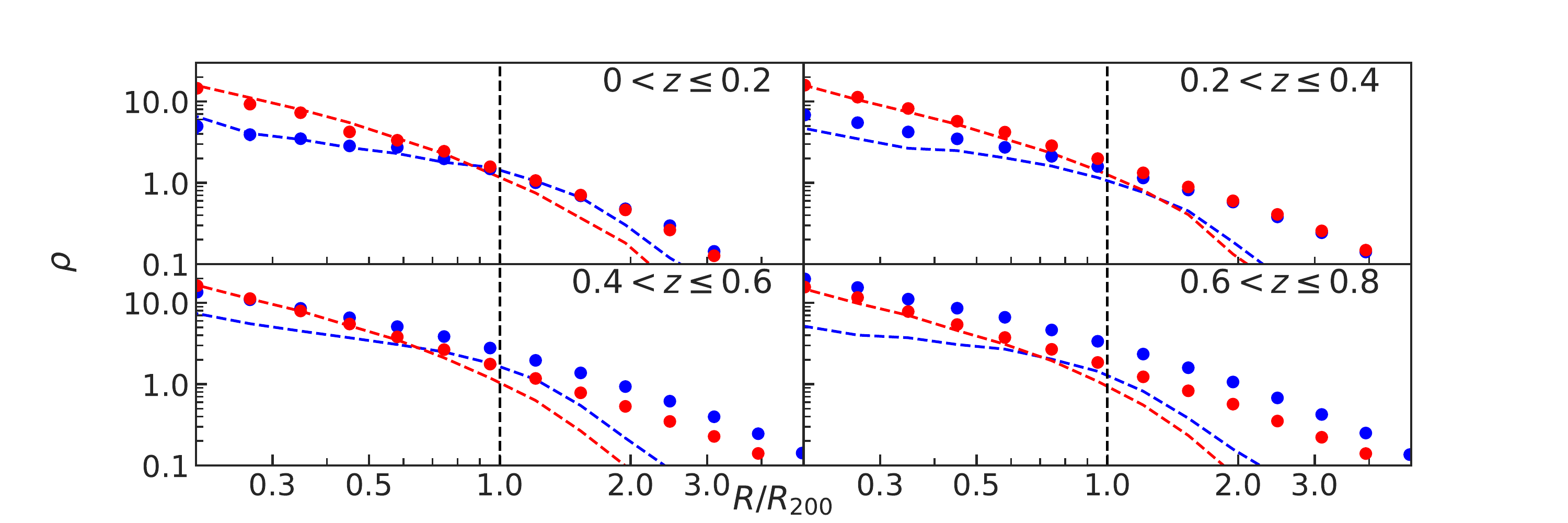}
	
	\caption{The stacked surface density ($\rho=N/{\rm Mpc}^2$) profiles of red and blue   galaxies  are displayed as a function of the distance from the cluster centre normalized to $R_{200}$, in different redshift bins. The dashed vertical line marks the position  that defines the red  fraction ($f_r$). Dashed curves show the profiles derived from the Illustris TNG300-1 models (see Section~\ref{sec:comp2TNG}). \label{fig:mfrac_zbin}}
\end{figure*}

\begin{figure*}
\subfloat[]{{\includegraphics[width=.5\linewidth]{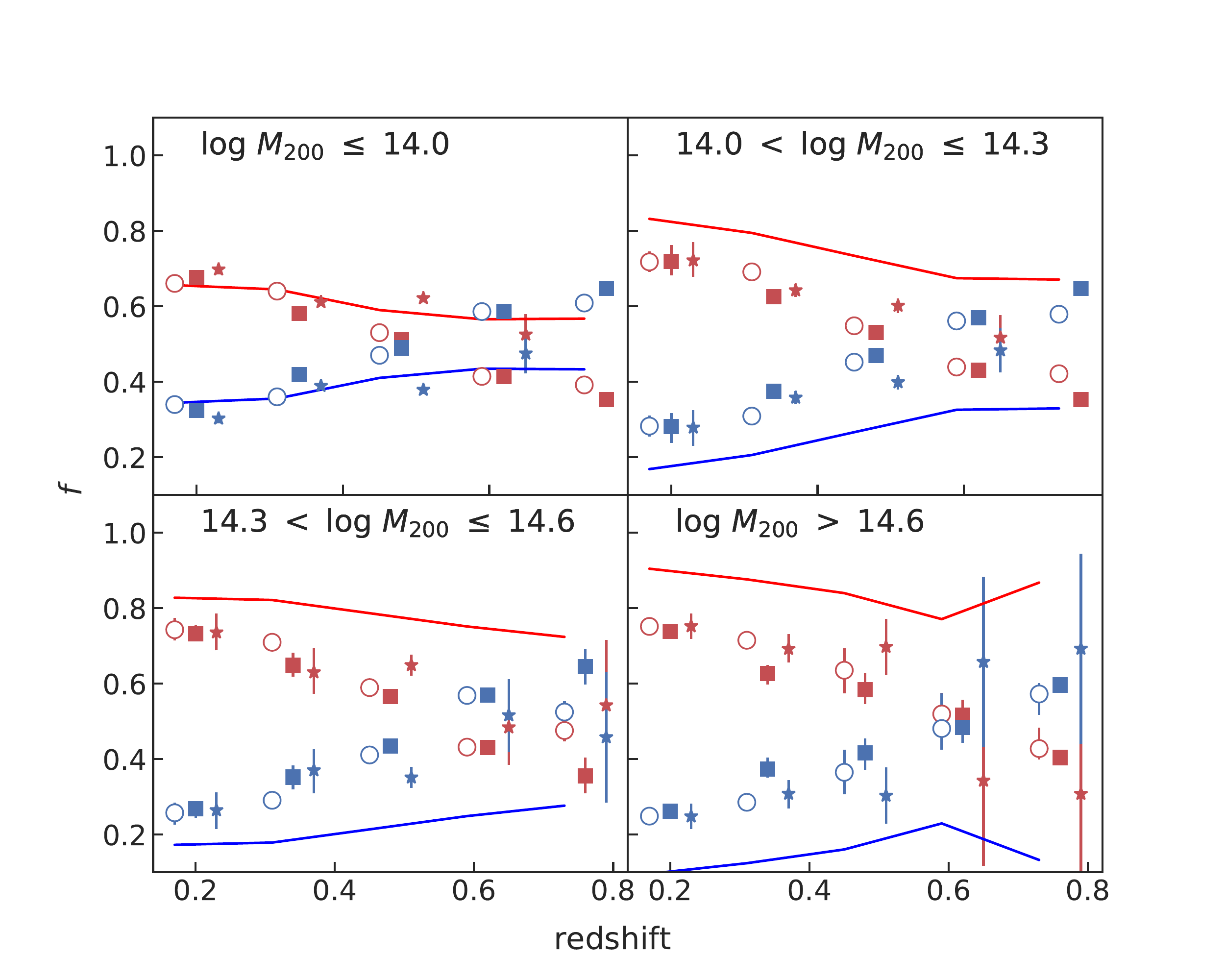} }}%
\subfloat[]{{\includegraphics[width=.5\linewidth]{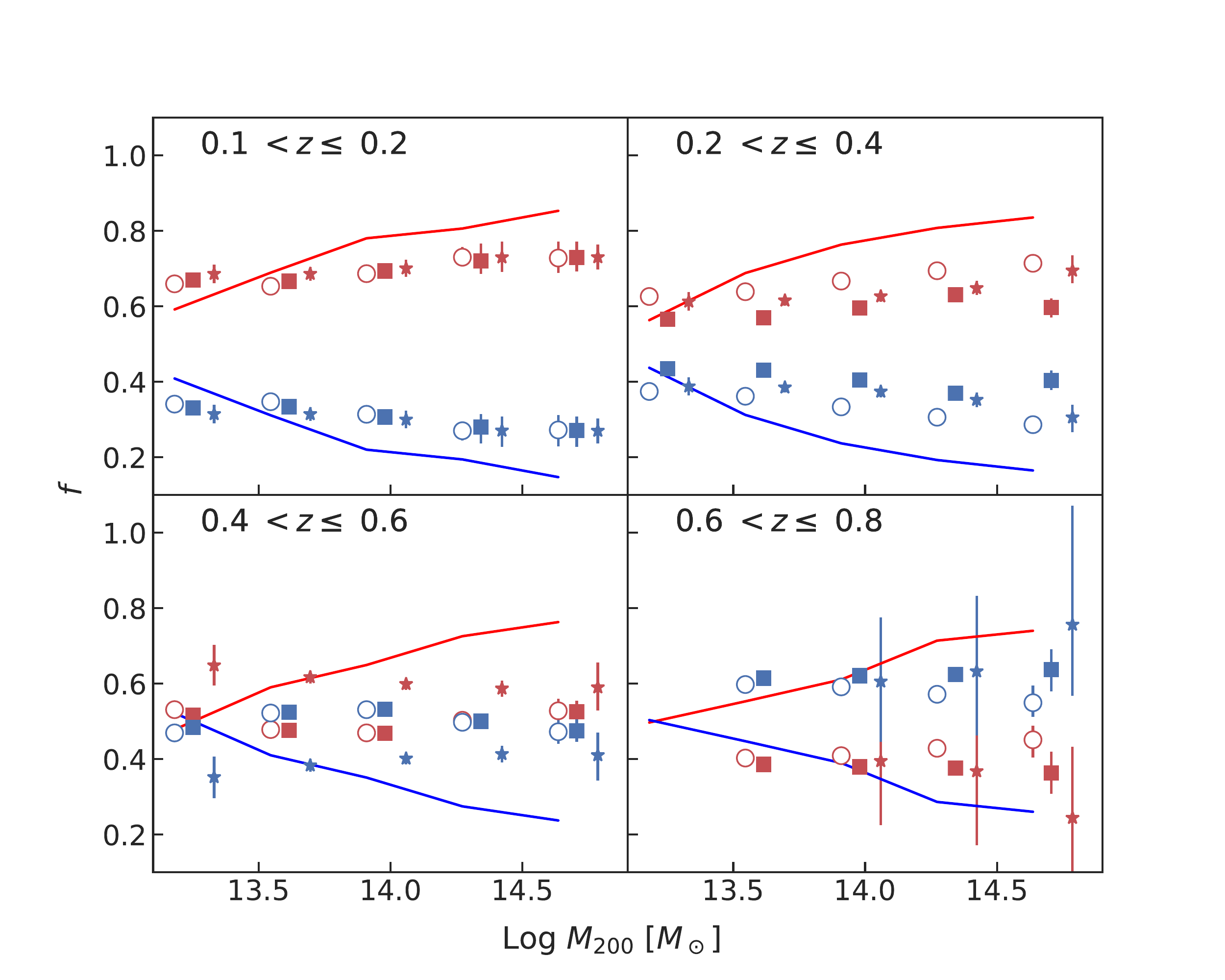} }}%

	\caption{Red and blue fractions of the galaxies are plotted in bins of ($a$) cluster mass and ($b$) redshift:  the fractions obtained using the colour, \lephare\  and structural parameter classifications are displayed by circles, squares and stars, respectively; an offset in the $x$-axis positions was applied to the \lephare\  and structural parameter symbols, for display purposes. The solid red and the blue lines show the values derived using the Illustris TNG300-1 simulations. \label{fig:clz_mass_type}}

\end{figure*}		

\section{Red and blue cluster galaxies}
\label{sec:red_blue}

Algorithms based on the detection of a red sequence like {\em redMaPPer} may introduce a bias in estimating the fraction of early- (red) and late- (blue) type galaxies, although they also include blue galaxies 
as cluster members \citep[see e.g.][]{2017MNRAS.467.4101G}.
Since \AMICO\  allows to search for clusters with no assumptions 
on the colours of their member galaxies, it reduces the risk of introducing such biases.
In this section we describe how blue and red cluster members are selected in the catalogue.

\subsection{Classification of red and blue galaxies}
Several methods were proposed by different authors to separate {\em red} and {\em blue} galaxies in clusters. 
A method often adopted  is to identify the cluster red sequence in the colour-magnitude diagram, and then select as red (blue) galaxies those whose colours are within (outside) a given distance in colour from the red sequence, for instance \citep[see e.g.][]{2011MNRAS.417.2817P}  $\Delta = [+0.1, -0.3]$ mag.
\citet{2006MNRAS.365..915A} pointed out that this selection introduces a bias  due to the evolving colours in redshift, so that at higher redshifts too many galaxies will be classified as {\em blue}: this may contribute to the so--called   
Butcher-Oemler effect \citep{1984ApJ...285..426B}, where clusters at increasing redshift are observed to present an increasing fraction of blue galaxies. 

To avoid this bias, \citet{2006MNRAS.365..915A} proposed two galaxy models based on the \citet{2003MNRAS.344.1000B}  population synthesis model with solar metallicity, a formation redshift $z_F = 11$ and an exponentially declining star formation with e-folding time $\tau=1$ and $\tau=3.7$ to describe early ({\em E}) and late ({\em Sa}) type  galaxies. They then defined as {\em red} ({\em blue}) galaxies those {\em redder} ({\em bluer}) than an {\em Sa} galaxy; an upper limit to the  color derived from the $E$ model allows to remove those galaxies that are too red at a given redshift to be likely cluster members. Here we adopt the same approach, using  \ezgal\ to derive the {\em E} and {\em Sa} models; the models are calibrated from observed galaxy colours at $z \sim 0.15$. To have a good separation both at lower and higher redshifts, we adopt as  colour ($\mathcal{C}$ hereafter) $g-r$  (if $z<0.4$) or $r-i$ (if $z>0.4$). We set the upper limit to $\mathcal{C}_E$ + 0.1 mag, to take into account the flux uncertainties and the scatter in the colour-magnitude relation due to  metallicity \citep[see e.g.][]{2019ApJ...870...70S}. Figure~\ref{fig:colBCG} compares the observed BCG and model colours: to visualize the expected scatter in colours due to metallicity, the model colours from two $E$ models at subsolar ($Z/Z_\odot$=0.5) and supersolar  ($Z/Z_\odot$=2) are displayed. 

As an alternative method,  \lephare\  
is used to separate the cluster members into red and blue galaxies \citep[see][for a similar approach]{2018A&A...613A..67S}. For each cluster, the redshift is fixed to the \AMICO\ cluster redshift, and \lephare\  is run to derive the best-fitting template: we use the \texttt{CE\_NEW} library in \lephare\ , which consists of  66 templates based on the \texttt{CWW}  \citep{1980ApJS...43..393C} SEDs. We define as {\em red} a galaxy best-fitted by a template describing Elliptical galaxies, as {\em blue} otherwise. To improve this morphological classification, we use the 9-bands optical + near--infrared photometry available in the KV450 release.

Finally, we also use structural parameters for a classification independent from the colours of the galaxies. From the catalogue of galaxies in KiDS-DR3 with reliable structural parameters ($\chi^2 < 1.3$, $0.5 < n_{S\acute{e}rsic} < 15$), we select  $\sim$ 580,000 cluster member galaxies: of these, 2583 are classified as BCGs.
Figure~\ref{fig:bcg_sersic} shows the distribution of the S\`ersic index ($n_{S\acute{e}rsic}$) and  axial ratio ($q$) for the cluster member galaxies classified as blue and red from their colour classification.   2424  are classified as red, 159 as blue, with average values of $n_{S\acute{e}rsic}$ =  $6.3\pm2.1$ and $1.4\pm1.2$ respectively.
For bright galaxies, $n_{S\acute{e}rsic}=2$ provides a good separation between red and blue galaxies \citep[see e.g.][]{2017MNRAS.472.2054P}. 
However, measurements of the structural parameters for faint and/or high redshift galaxies are hampered by an increasing uncertainty. For this reason in the following discussion on red/blue cluster members we use these results only as a comparison based on a completely independent approach.

\subsection{Cluster red sequence}
The membership probability and blue/red classification are used together to fit the cluster red sequence. To this end we select galaxies classified as early-type, with a probability membership $>$ 50\%. We further select the galaxies in the magnitude range $r_{\rm BCG} + 0.5 < r < r^*+2$ to remove the BCGs, which  may significantly deviate from the colour-magnitude relation with respect to other cluster galaxies \citep[see e.g.][]{2009MNRAS.394.2098S}. The fit is done with a robust regression where the Tukey's Biweight function is used as the M-estimator \citep{Venables}. For both $g-r$ and $r-i$ vs. $r$ (observed colours) we are not able to  detect any variation of the slope with redshift: for $g-r$, we obtain a slope of $-0.04 \pm 0.04$ in the redshift bin $0<z<0.4$ and  $-0.04 \pm 0.08$ in the redshift bin $0.4<z<0.8$; for $r-i$, the slope is $-0.01\pm0.01$ ($0<z\le0.4$) and $-0.02\pm0.04$ ($0.4<z<0.8$).

\subsection{Density profiles of blue and red members}

For each cluster, a surface number density profile is derived  counting  the red and blue members with an $r-$band magnitude  brighter than $m_*+1.5$, $m_*$ being the characteristic magnitude defined in Section~\ref{sec:bcg_selection}; each galaxy is weighted by its membership probability, and the resulting number divided by the area of the bin. 

To analyze the radial dependence of the density of the red and blue cluster members,  stacked surface density profiles are derived by summing the density profiles in different clusters and normalizing over the total number of clusters.
The radial profile of the red and blue densities  for four redshift bins with a width of $\Delta z=0.2$ and centred at $z=[0.1,0.3,0.5,0.7]$ is presented  in  Figure~\ref{fig:mfrac_zbin}. Here, the distance of the galaxies from the cluster centre is normalized by $R_{200}$, derived from the \AMICO\ cluster mass ($M_{200}$, see Eq.~\ref{eq:mscale}) based on a Navarro-Frenk-White \citep{1997ApJ...490..493N} profile.
Within $R_{200}$ the density is larger for red than for blue members: the red/blue ratio decreases with increasing distance from the cluster centre. This is in agreement  with what found e.g. by \citet{2017MNRAS.467.4015H}, \citet{2018PASJ...70S..24N}: the value of the cluster red fraction  strongly depends on the  distance from the cluster centre   \citep[see also][]{2018MNRAS.481.4158W}. We hereafter define as the {\em red cluster fraction} ($f_r$) the density ratio of red and blue cluster members within $R_{200}$.

Figure~\ref{fig:clz_mass_type} displays this fraction for different cluster mass and redshift bins. The same analysis is applied to the blue and red classification derived using colours, the \lephare\  best-fit templates and the structural parameters; only clusters with a minimum of 5 member galaxies are selected for this analysis.  All classification methods show that the red fraction is  $\sim$ 70\% at low redshifts ($z\sim 0.2$) and decreases ($\sim$ 50\%) as the redshift increases, in agreement with what found in other cluster studies \citep[e.g.][]{2017MNRAS.467.4015H, 2018A&A...613A..67S, 2018MNRAS.481.4158W}.  
 At redshift $z>0.4$, the colour and \lephare\  classifications indicate that the red fraction starts to be lower than the blue fraction. In the case of the structural parameter classification, the red fraction still significantly decreases, but not below the blue fraction. The uncertainties are however large, due to the low number of  galaxies with meaningful structural parameter measurements: there are  $\sim$ 40\,000 galaxies within $R_{200}$, compared to the over 450\,000 available with the classification based on colours.  At redshifts $z>0.6$ the results are 
displayed, but the uncertainties are even larger.
 The red fraction increases for increasing cluster masses, with a steeper increase when $z<0.4$ and $\log M_{200} < 14$ 
\citep[see][for a similar result]{2009ApJ...699.1333H, 2018A&A...613A..67S}. At higher redshift,  the uncertainties do not allow do draw definite conclusions.

To further verify that there is no systematic effect due to how blue/red galaxies are defined, we select a sample of  galaxies with a low membership probability ($p_{\rm memb}$ < 20\%): these are more  
likely to be field galaxies rather then cluster members. 
For this sample, we find that the  red fraction is $<$ 0.4 at all redshifts, in agreement with what found by \citet{2017MNRAS.472.2054P}.

\begin{figure}
	\includegraphics[clip,width=1.0\linewidth]{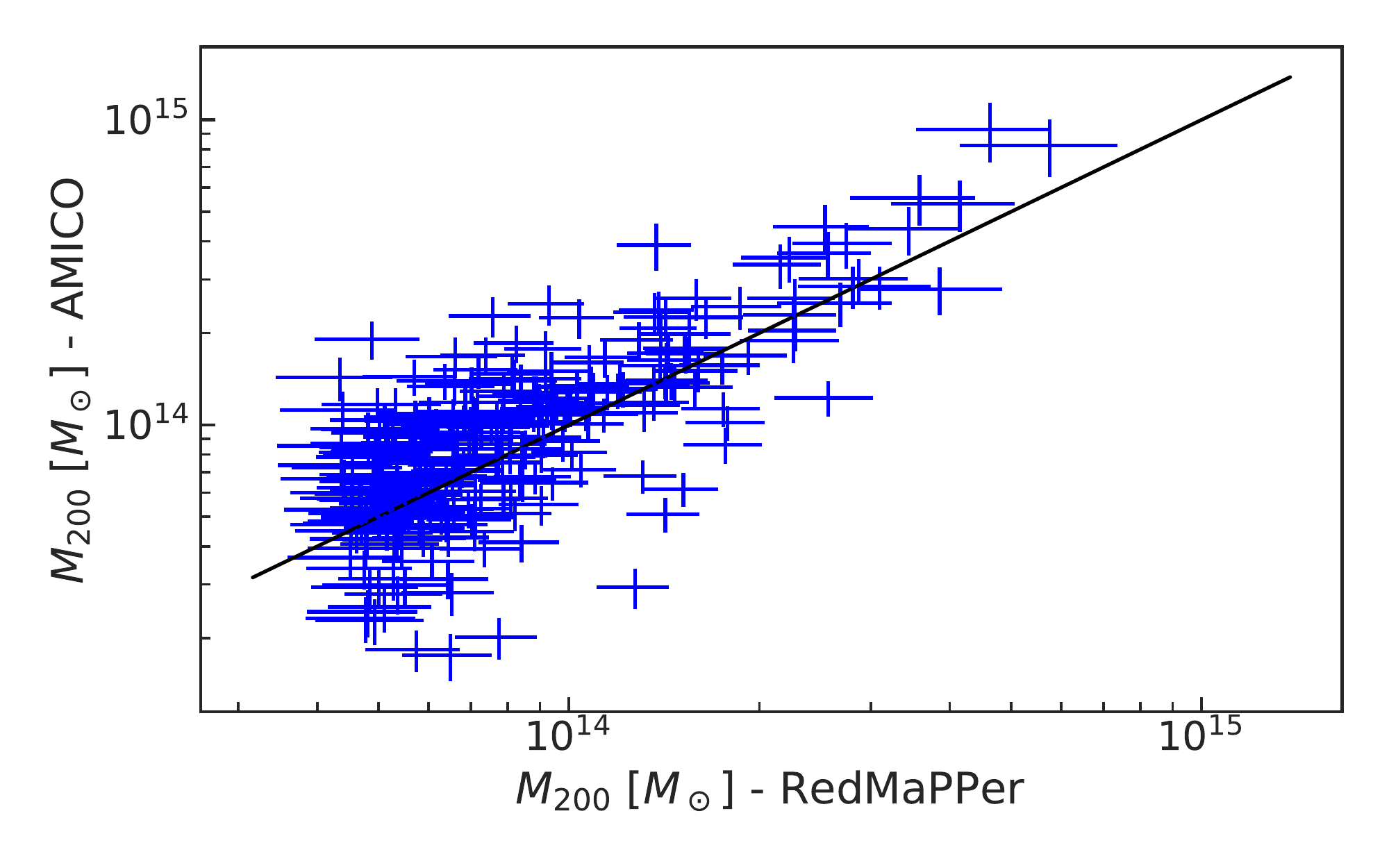}
	
	\caption{ Comparison between masses for matched clusters in the KiDS-\AMICO\ and {\em redMaPPer} catalogues, in the redshift range $0.1 < z < 0.33$ where the {\em redMaPPer} mass calibration was done. The solid line shows the 1-to-1 relation as reference.
		\label{fig:m200_k2rm}}
\end{figure}

\begin{figure}
	\includegraphics[clip,width=1.0\linewidth]{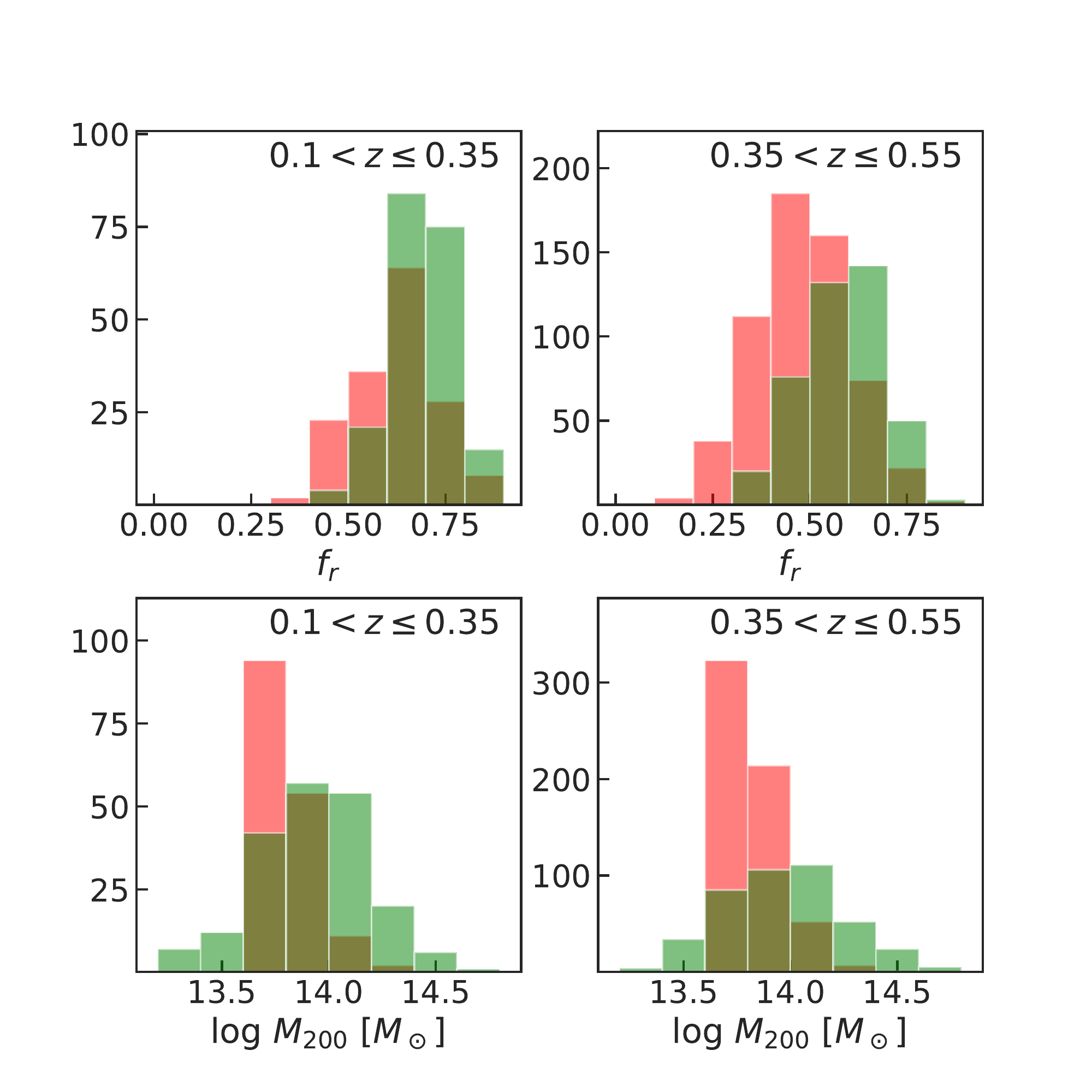}
	
	\caption{Distribution in fraction of red galaxies ($f_r$) and mass (from \AMICO) in two redshift bins for the matched (green) and unmatched (red) clusters from the KiDS-\AMICO\ vs. {\em redMaPPer} catalogues. A cut on mass, $\log M_{200} > 13.5 M_\odot$, was applied to unmatched clusters. 
		\label{fig:blueredz_a2r}}
\end{figure}

\begin{figure}
	\includegraphics[clip,width=1.0\linewidth]{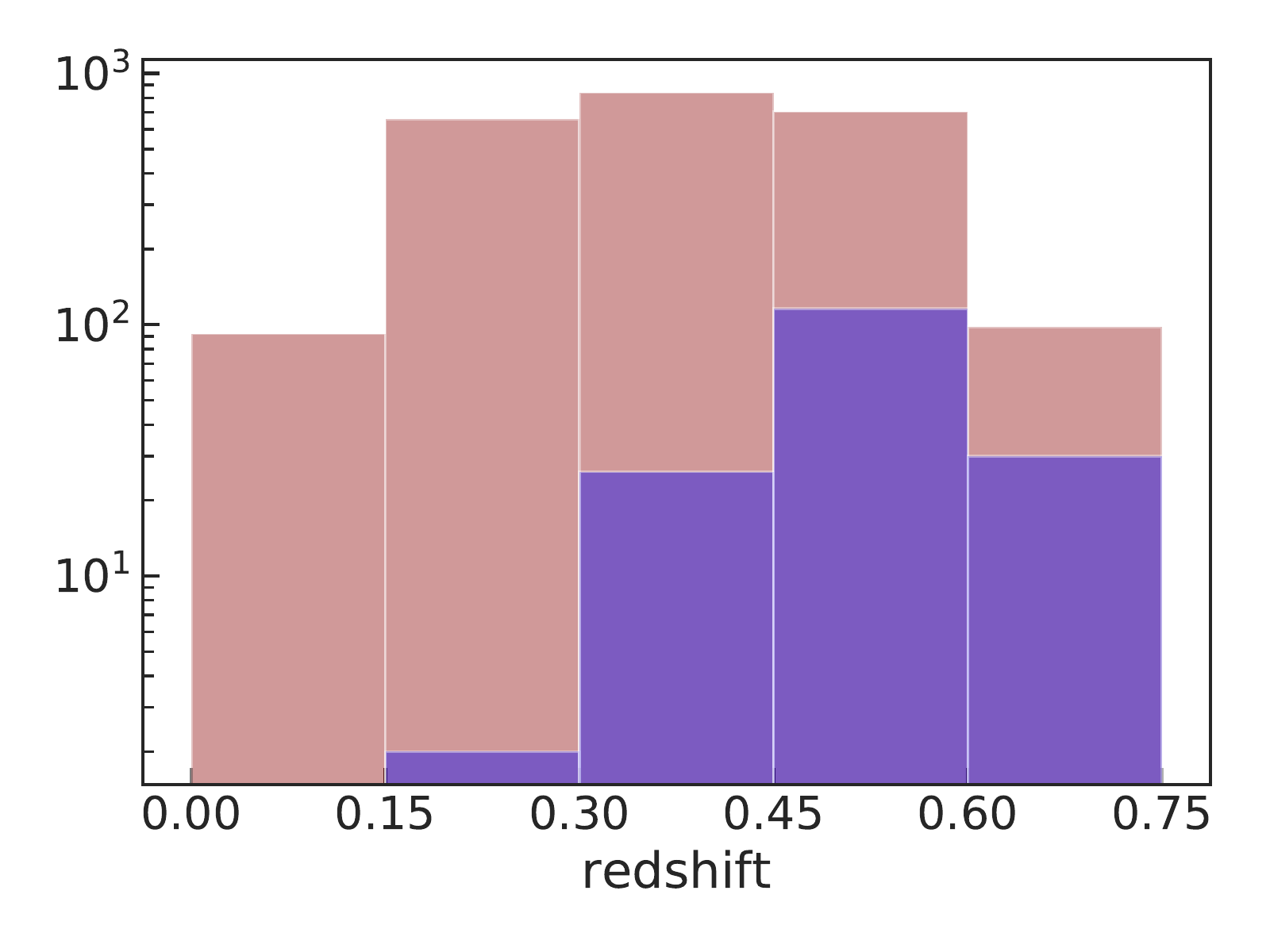}
	\caption{Redshift distribution of red and blue  BCGs.
		\label{fig:zdistBCG}}
\end{figure}

\subsection{Comparison with the  SDSS-redMaPPer catalogue}
\label{sec:comp2RM}

\citet{2019MNRAS.485..498M} compared the KiDS-\AMICO\ DR3 and the SDSS {\em redMaPPer}  \citep[v. 6.3,][]{2014ApJ...785..104R, 2015MNRAS.453...38R}  cluster catalogues in the redshift range 0.08 $<z<$ 0.55. They found that 624 (92\%) of the 681 clusters detected by {\em redMaPPer} in the common area (236 sq. degrees in KiDS-N) were matched by detections in the KiDS-\AMICO\ catalogue; conversely, 3498 clusters are detected by KiDS-\AMICO\ which are not found by {\em redMaPPer}.  To verify if the unmatched clusters reflect some intrinsic differences in their properties,  we  first need to account for the different richness cuts by the two catalogues. Since these are defined in different ways in the two algorithms (SN$>$3.5 and $\Lambda_{RM}>20$ for KiDS-\AMICO\ and {\em redMaPPer} respectively), we  use the mass derived with the respective scaling relations \citep{2019MNRAS.484.1598B,2017MNRAS.466.3103S}. 
In {\em redMaPPer} the calibration was done using clusters in the redshift range $0.1 < z < 0.33$, and masses were measured within the radius where the mean density is 200 times the mean matter density of the Universe at that redshift ($M_{200m}$), rather than the critical density as in \AMICO\ ($M_{200c}$). The conversion from  $M_{200m}$  to  $M_{200c}$ is done assuming a concentration $c_{200}=4$ \citep{2019MNRAS.484.1598B}. 
The masses of matched clusters are displayed in Figure~\ref{fig:m200_k2rm}, showing the good agreement between the two estimates, with a median offset $\log M_{200,RM}-\log M_{200,AMICO} \sim 0.05$.
We finally select only unmatched KiDS-\AMICO\ clusters with $M_{200} > 10^{13.5} M_\odot$, the lower mass limit in {\em redMaPPer}, that gives  $\sim$ 1200 unmatched clusters.
 Figure~\ref{fig:blueredz_a2r} compares the distribution in matched and unmatched clusters of the red fractions ($f_r$) defined in Sec.~\ref{sec:red_blue}, in two redshift bins: $0.08 < z < 0.35$, where the {\em redMaPPer} masses were calibrated,  and  $0.35 < z < 0.55$, where instead the {\em redMaPPer} masses were extrapolated. There are more unmatched clusters in the high redshift bin, with lower red fractions ($f_r < 0.6$) than at lower redshifts. 
It should be however noted that 90\% of the unmatched clusters have $M_{200} < 10^{14}  M_\odot$: due to uncertainties related to  the richness estimate and the mass scaling relations, we are not able to conclude if the unmatched systems are not in 
{\em redMaPPer} because of the red-sequence selection on which the method stands, or because their mass is overestimated (underestimated) by the KiDS-\AMICO\ ({\em redMaPPer})  calibration.

The {\em redMaPPer} catalogue also lists the 5 central galaxies, and the one adopted as the cluster centre. For the common clusters, we find that for 79\% of them the BCG  selected in this paper  is one of the  {\em redMaPPer} 5 central galaxies: of these, only 2\% are classified here as {\em blue} BCG. Of the remaining BCGs which are {\em not} one of the  {\em redMaPPer} 5 central galaxies, 87\% are red and 13\% are blue.

\begin{figure}
	\includegraphics[clip,width=1\linewidth]{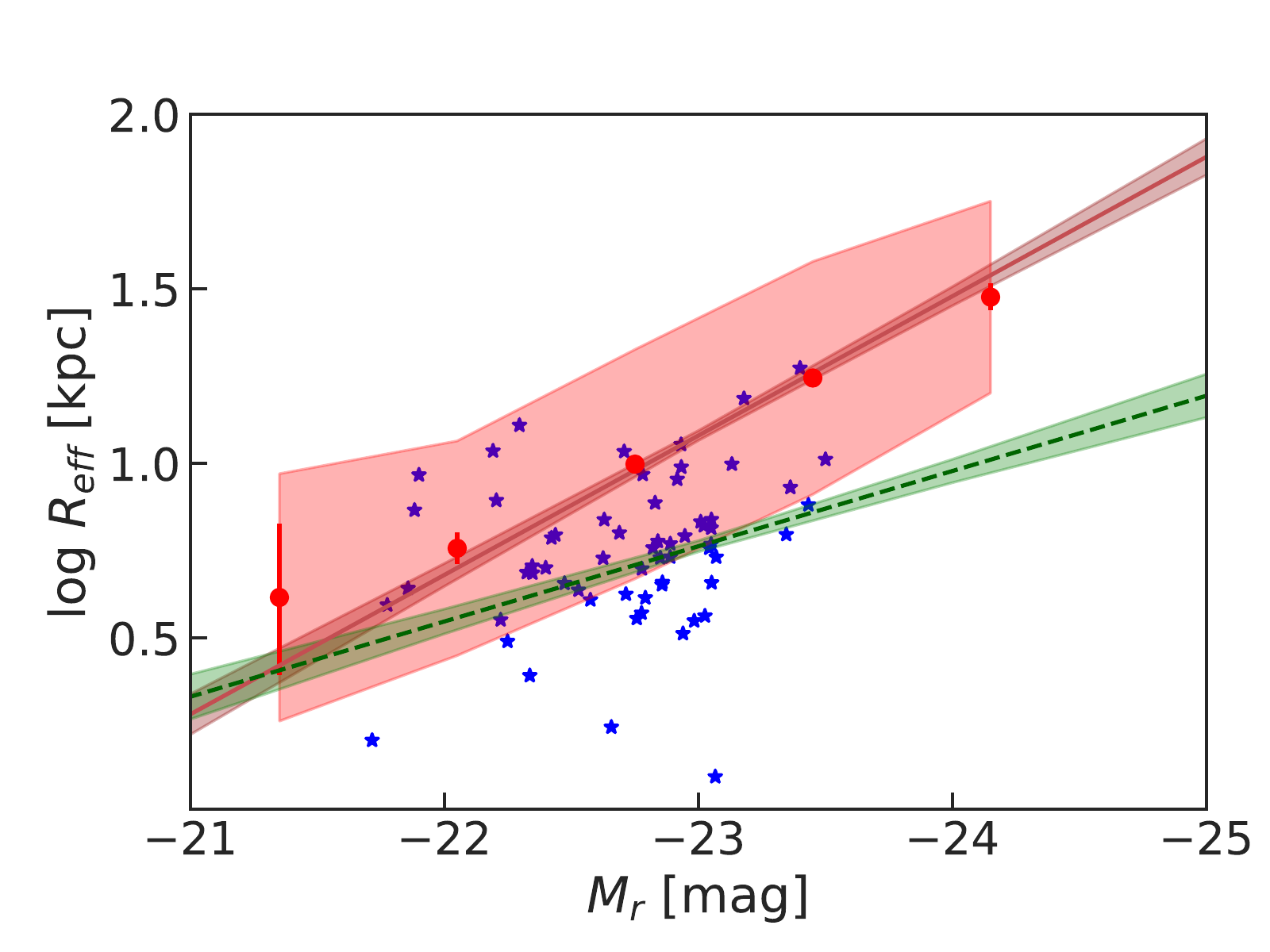}
	\includegraphics[clip,width=1\linewidth]{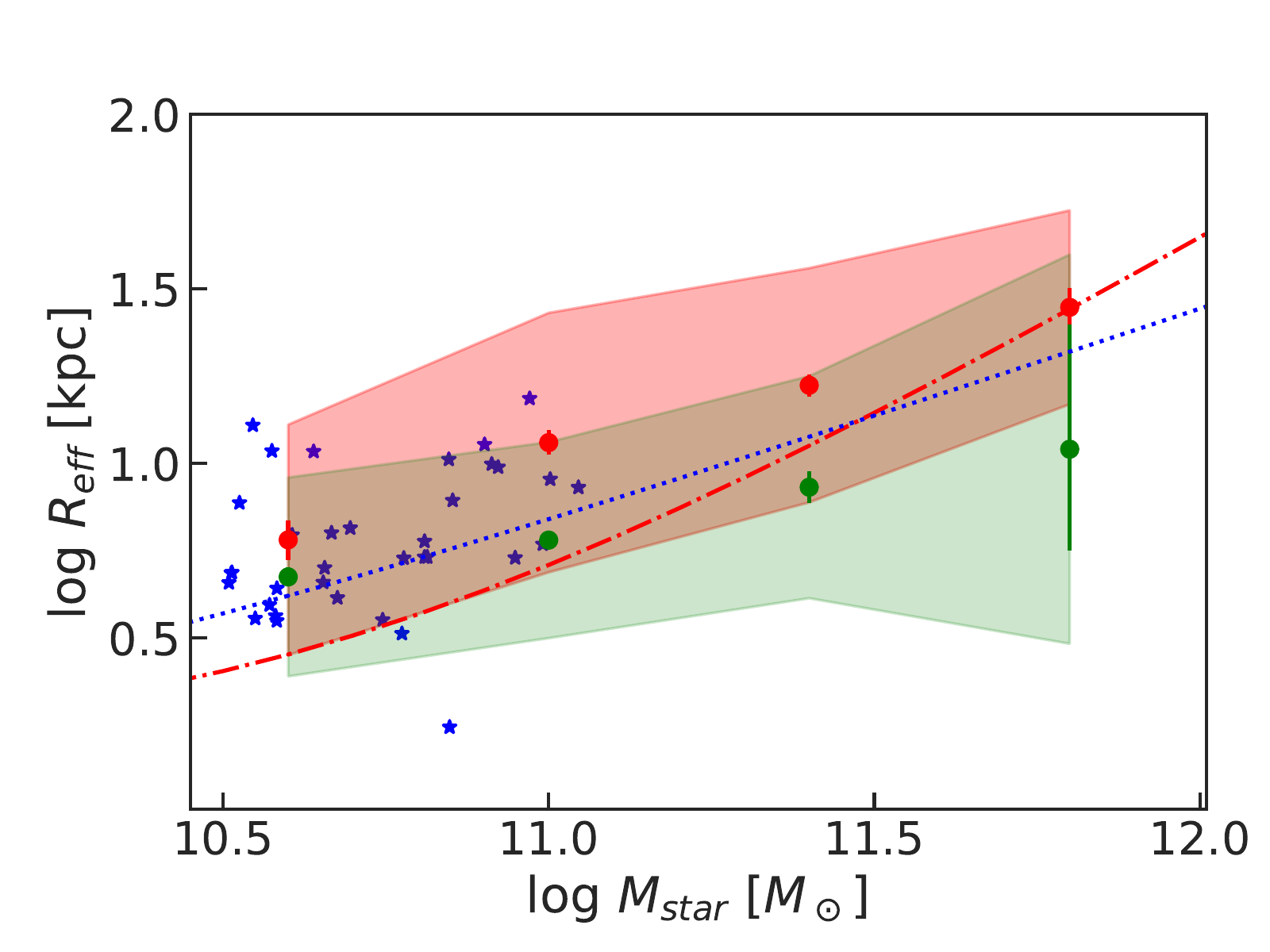}
\caption{{\em Top:}  size-luminosity distribution in  BCGs. 
The wide area (light red) shows the 1-$\sigma$ scatter for observed values, whose bootstrapped average values and error bars, and linear fits, are plotted in red. Values for blue BCGs are displayed as blue stars. The linear fits for  galaxies with a low membership probability are displayed in green; shaded areas display 95\% confidence intervals.   {\em Bottom}: 
the size-stellar mass is displayed for BCGs with $0.3 < z < 0.5$  (same symbols as above) and compared with the fits at $z\sim0.4$ derived by  \citet{2018MNRAS.480.1057R} for spheroid (dashed--dotted red) and disc (dotted blue) galaxies. The wide green area shows the 1--$\sigma$ scatter for low membership  probability galaxies. }
		\label{fig:rabs_reff}
\end{figure}

\begin{figure*}
	\includegraphics[clip,width=1.02\linewidth]{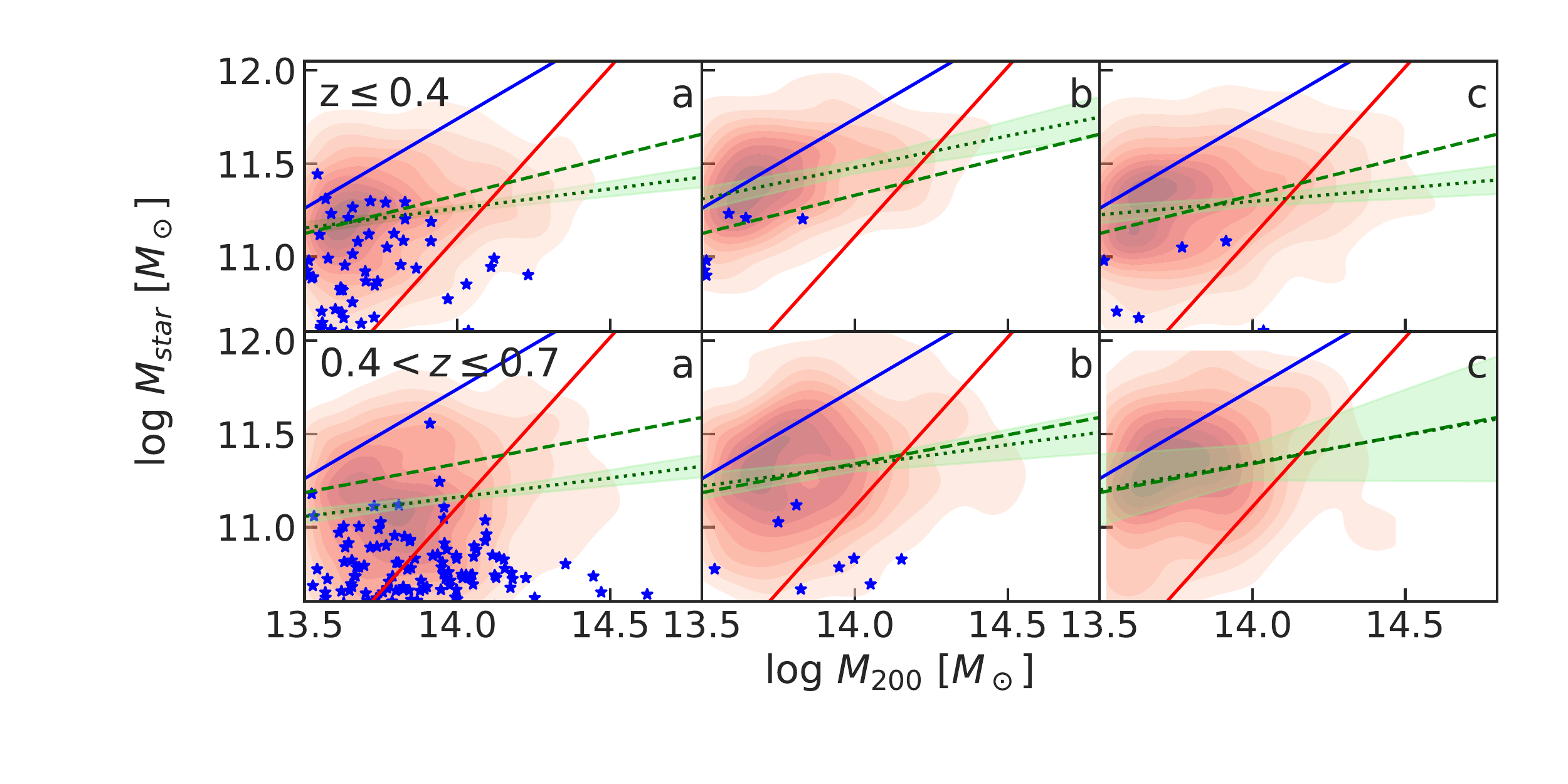}
	\includegraphics[clip,width=1.02\linewidth]{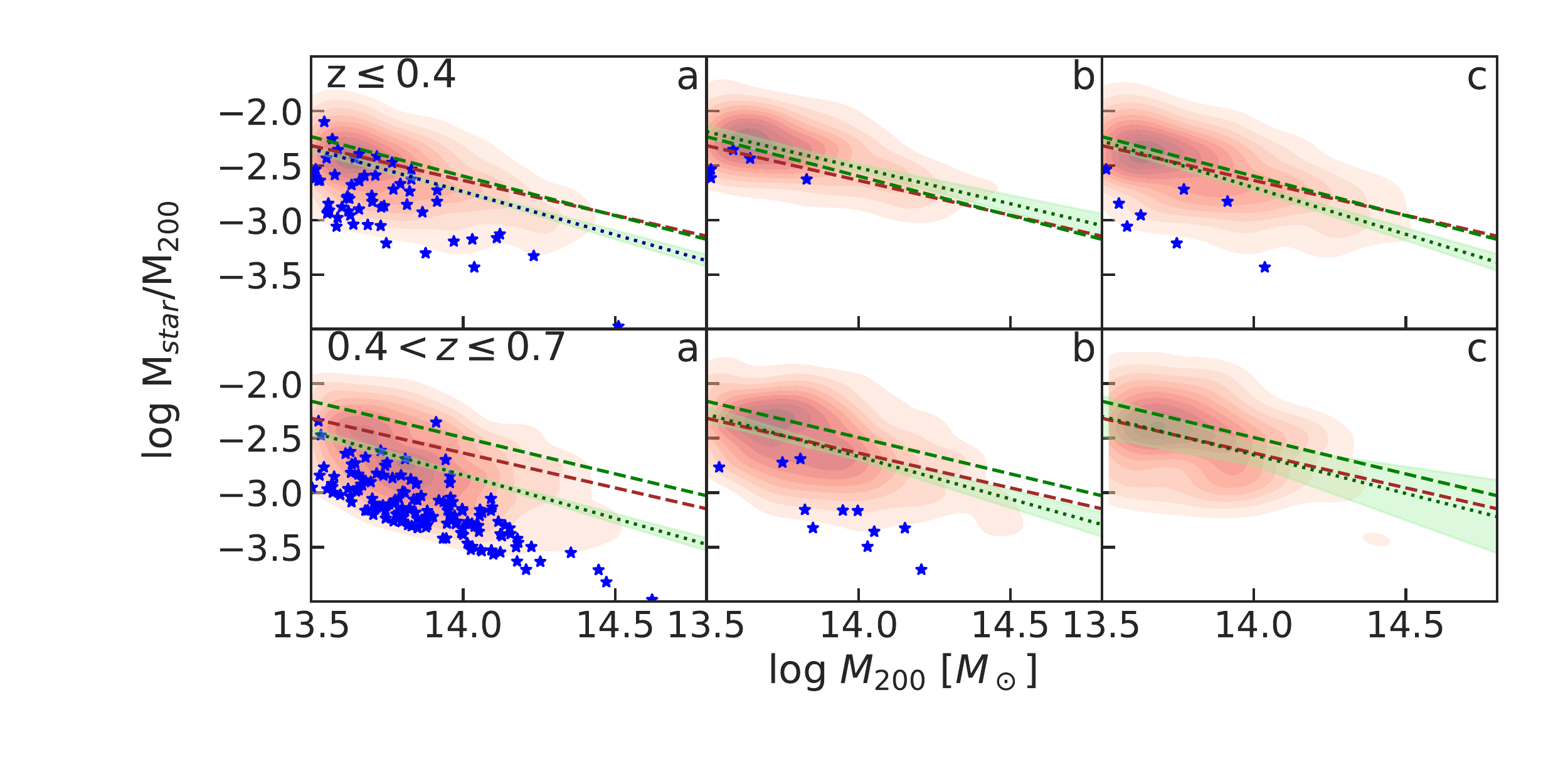}
	
	\caption{
	The plots show the distribution of the BCG stellar mass and stellar to cluster mass ratio vs. the  cluster mass (red contours), with blue BCGs plotted as dark-blue 
	stars.  
	The upper/lower rows show clusters  with redshift $z\le0.4$ and $0.4 < z \le 0.7$ respectively,
	with (from left to right): ($a$) no selection, ($b$) $\Delta M_{1,2} > 1$, ($c$) $f_r>0.7$. 
	The relations fitted from the data and 95\% confidence levels are displayed in green.
	In the {\em upper} plots, the blue and red lines display fits derived by \citet{2016MNRAS.462.4141L}, weighted to relaxed and unrelaxed  clusters respectively. The dashed green line shows the fit from \citet{2019A&A...631A.175E}.
	In the {\em bottom} plots, the red and green lines are the fits derived by \citet{2013MNRAS.428.3121M} and 
	\citet{2020A&A...634A.135G} respectively. 
     \label{fig:mbcg} 
		}
\end{figure*}

\begin{table}
\centering
\caption{Coefficients of the fits  ($y = mx + c$) for the $\log M_{\rm star}$ vs. $\log M_{200}$ relation, with the different selections ($a,b,c$) described  in Figure~\ref{fig:mbcg}.}
\label{tab:mstar_m200}
\begin{tabular}{lccc}
\hline
  & m & c & N \\
\hline

\multicolumn{3}{l}{$0.1<z\leq 0.4$}\\

 a & 0.21 $\pm$ 0.03 &8.32 $\pm$ 0.38 &1551 \\ 
 b & 0.34 $\pm$ 0.05 &6.75 $\pm$ 0.73 &329 \\ 
 c & 0.14 $\pm$ 0.04 &9.29 $\pm$ 0.54 &620 \\ 
\multicolumn{3}{l}{$0.4<z\le 0.7$}\\
 a & 0.21 $\pm$ 0.03 &8.26 $\pm$ 0.43 &1789 \\ 
 b & 0.22 $\pm$ 0.06 &8.23 $\pm$ 0.81 &460 \\ 
 c & 0.29 $\pm$ 0.17 &7.25 $\pm$ 2.30 &71 \\ 

\hline
 \end{tabular}
\end{table}

\section{BCG properties}	
\label{sec:other}

\subsection{Blue BCGs}

We find that  $\sim$ 7\% of the BCGs are not classified as red galaxies: this fraction is consistent with what found e.g. in the SDSS  \citep{2011MNRAS.417.2817P,2019MNRAS.487.3759C}.  The redshift distribution for the BCG classified as 'red' and 'blue' is displayed in Figure~\ref{fig:zdistBCG}: the fraction of blue/red BCGs is $<$5\% at $z<0.3$ and increases to $>$10\% at $z > $0.4, consistently with \citet{2011MNRAS.417.2817P}.

\subsection{Size and luminosity}

To further analyze the BCG properties (BCG stellar mass, size and luminosity) and compare them with the cluster mass ($M_{\rm 200}$), we select a subsample of 2000 BCGs (of which 60 blue) with  $M_{\rm star}/M_\odot> 10^{10.5} M_\odot$ for which measurements of the size  are available.  The size is the effective radius produced by the fit with the S\'ersic profile (see Section~\ref{sec:data}). The  BCG luminosity is the $k$-corrected $r-$band absolute magnitude, where the $k-$correction terms are derived from the \lephare\  output.

Figure~\ref{fig:rabs_reff} ({\em top}) shows that  there is a correlation between the size and luminosity ($L \propto 10^{-0.4 M_r}$) of the BCGs:  this is confirmed by  the Spearman test, giving a coefficient of 0.5, with a null probability of no correlation. 
At a given luminosity, red galaxies  with a low ($<$ 20\%) membership  probability have a significant
smaller size than BCGs at the same luminosity:  since  field early-type galaxies are expected to have a smaller size than 
BCGs in clusters \citep{2007MNRAS.379..867V,2007AJ....133.1741B}, this agrees with a higher contamination from non-cluster members as the membership probability decreases. Red cluster members selected to have a more significant membership probability ($>$ 50\%), but fainter than the BCGs
($r > r_{\rm BCG} + 0.5$) are in an intermediate position. To summarize, 
we obtain $R_{\rm eff} \propto L^{0.96\pm0.04}$ (red BCGs), $R_{\rm eff} \propto L^{0.87\pm0.03}$ (non-BCG red cluster members) and $R_{\rm eff} \propto L^{0.53\pm0.05}$ (red galaxies with low membership probability). A 
size-luminosity relation steeper in BCGs than in the  bulk of early-type galaxies  ($r_{\rm BCG} \propto L_{\rm BCG}^{0.88}$ and $r \propto L^{0.68}$ respectively) was reported by \citet{2007AJ....133.1741B}. As it concerns blue BCGs, we are not able to derive a significant fit due to their lower number and larger scatter; however, compared to red BCGs  their size-luminosity relation appears to be more similar to what measured in galaxies with low membership probability.

The existence of a difference in the stellar mass-size relation in clusters with respect to field galaxies, though predicted by hydrodynamical models,  is controversial: for instance \citet{2013MNRAS.428.1715H, 2013ApJ...779...29H} did not find any evidence for differences in the mass-size relation among field, group and cluster galaxies, which was instead detected by \citet{2018MNRAS.480..521H}. Figure~\ref{fig:rabs_reff} ({\em bottom}) compares the values for BCGs in our catalog with the fits derived by \citet{2018MNRAS.480.1057R} for spheroid and disc  galaxies in KiDS without any selection on the environment, at redshifts $0.3< z< 0.5$ (similar results are obtained at lower redshifts). The size of red BCGs appears to be larger at a given mass than for other galaxies, though the scatter is large. For stellar masses $M_{\rm star} > 10^{11.5} M_\odot$, the difference is lower, but the sample on which the \citet{2018MNRAS.480.1057R}  fits were derived may be more contaminated by cluster galaxies than at lower stellar masses. Blue BCGs instead, consistently with what already described above, are much closer to the values observed in other galaxies.

\begin{figure}
	\includegraphics[clip,width=1.02\linewidth]{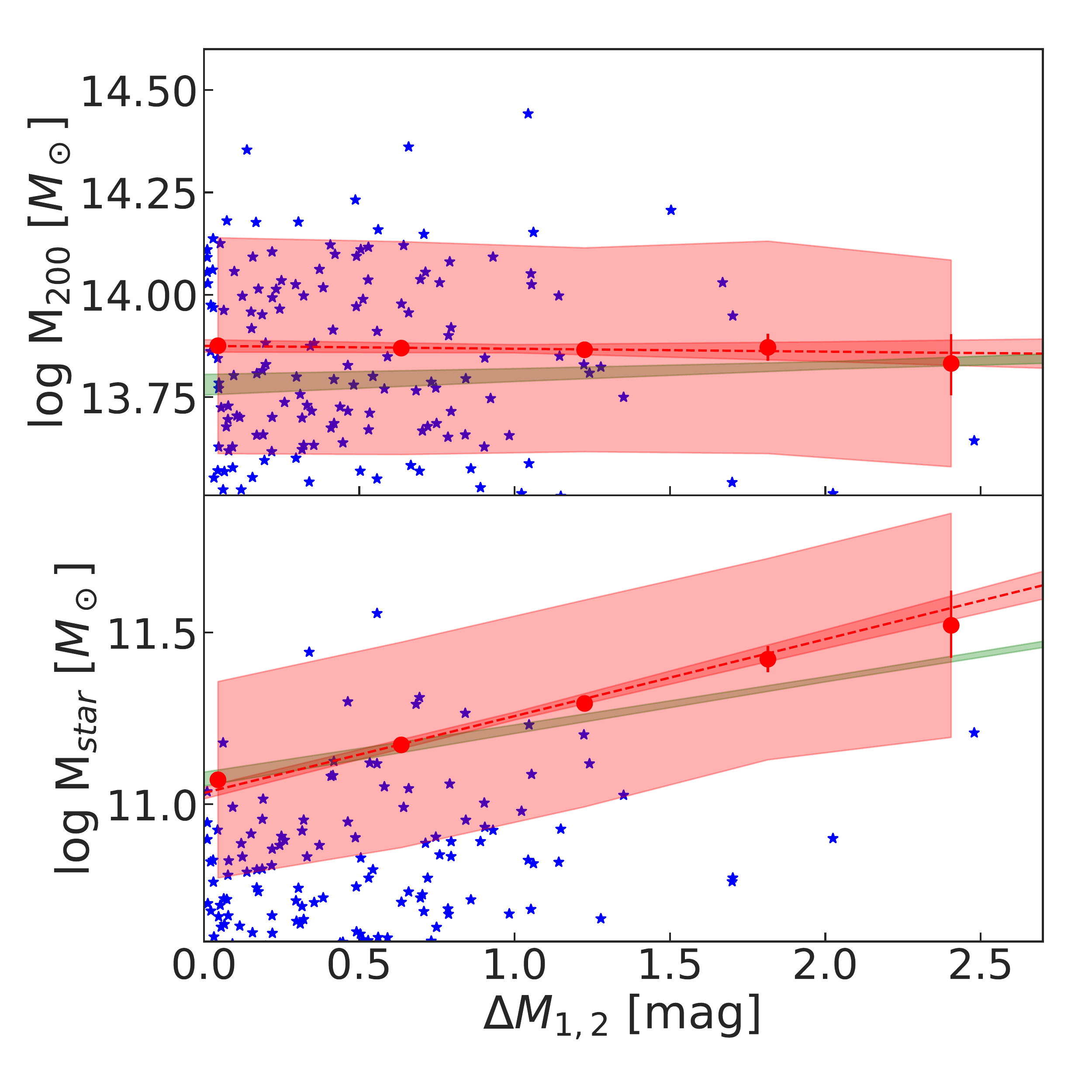}
	\caption{Total cluster mass (top panel) and stellar
mass (bottom) as a function of the magnitude gap. The wider area (light red) shows the 1-$\sigma$ scatter for observed values, whose bootstrapped average values and error bars are plotted as red circles. Blue BCGs are displayed as blue stars.  The fits derived from the Illustris TNG300-1 simulations are marked in green. Shaded areas display 95\% confidence intervals.
		\label{fig:mgap2Mstar}}
\end{figure}

\subsection{Cluster mass vs. BCG stellar mass, and dynamical status}

The connection between the BCG stellar mass and the cluster mass is predicted by hierarchical cluster formation models \citep{2013MNRAS.430.1238H,2018ApJ...860....2G,2019ApJ...878...14G,2020MNRAS.493.1361F}.
We therefore expect to see a  correlation \citep[see e.g.][and references therein]{2019A&A...631A.175E} between the cluster mass and the BCG stellar mass, or the stellar  to halo mass ratio (SHMR).

Figure~\ref{fig:mbcg} (top)  shows the distribution of $\log M_{200}$ vs. $\log M_{\rm star}$ for clusters in the two redshift bins ($z\le0.4$ and $0.4 < z \le 0.7$): in both bins, we find  a linear correlation between $\log M_{200}$ and  $\log M_{\rm star}$, with a moderate correlation cofficient $< 0.2$ given by  the Spearman test, and a null probability of no correlation, but with a large scatter. \citet{2016MNRAS.462.4141L} showed that the  relation between the cluster and BCG stellar mass  depends on the dynamical status of the clusters (relaxed and non--relaxed): part of the scatter that we observe could be therefore due to the fact that our sample includes clusters in a wide range of conditions.
In particular in the lower redshift bin, there is a good agreeement between the observed SHMR and the predictions from the models from \citet{2013MNRAS.428.3121M,2020A&A...634A.135G}, which are  derived by
sub-halo abundance matching models with  two  different stellar mass functions based on SDSS and COSMOS data respectively. In our data, at a given cluster mass blue BCGs show lower  stellar masses than red BCGs.
A similar trend was described by \citet{2019MNRAS.487.3759C}, who found a strong decline of the fraction of star-forming BCGs  in their sample both with the stellar and cluster mass.

To verify if there is any dependence on the cluster sample selection, in Figure~\ref{fig:mbcg}, panels $b$ and $c$, we show the results of the same analysis performed on clusters  with:  a magnitude gap, that  is the luminosity difference between the two brightest galaxies in the cluster\footnote{The magnitude gap is defined here as $\Delta M_{1,2} = M_{r,2} - M_{r,1}$, where $M_{r,i}$ is the $k$-corrected $r$-band absolute magnitude of the $i$-th brightest galaxy.}, $\Delta M_{1,2} > 1.0$; a red fraction $f_r > 0.7$, to select only clusters 
where the component of red galaxies is dominant as observed in the low-redshift clusters. The magnitude gap can be used as an 
indicator of the cluster dynamical status \citep[see e.g.][]{2013MNRAS.436..275W,2018MNRAS.478.5473L}, to separate relaxed  
($\Delta M_{1,2} > 1.0$) and 
disturbed ($\Delta M_{1,2} < 1.0$) clusters \citep{2018MNRAS.478.5473L}. Consistently, Figure~\ref{fig:mbcg} shows that  clusters with 
$\Delta M_{1,2} > 1.0$ are closer to the position expected from  the \citet{2016MNRAS.462.4141L} fits 
for relaxed clusters, both in the low and high redshift bins. 
Almost no blue BCG with $\Delta M_{1,2} > 1.0$ is observed. In the high redshift bin, selecting clusters with a low fraction of blue galaxies also  moves clusters to the position of relaxed clusters. 
To have a more quantitative estimate of these dependencies, we show in Table~\ref{tab:mstar_m200} the robust linear fits obtained for each selection, which may be compared 
e.g. with what obtained by \citet{2019A&A...631A.175E} for a sample of X--ray selected clusters. Both in the low and high redshift bins there is a better agreement with their parameters in the panels denoted with $b$, suggesting that the dynamical status has indeed  to be taken into account.

Figure~\ref{fig:mgap2Mstar} compares the magnitude gap with the cluster mass and the  BCG  stellar mass. The Spearman test gives a correlation coefficient of $\sim$ 0.4 for the BCG stellar mass, with a null probability of no correlation: systems with larger magnitude gap tend to show a BCG which is  more massive. Instead, we see  no significant correlation of the magnitude gap with the cluster mass. The same trend is seen in the Illustris TNG300-1 simulations, which are discussed in detail in Section~\ref{sec:comp2TNG}.
As it concerns blue BCGs, consistently with what discussed above they show  lower stellar masses than red BCGs, and a magnitude gap  $\Delta M_{1,2} < 1.0$.

\section{Comparison with Illustris-TNG simulations}
\label{sec:comp2TNG}

The Illustris TNG300-1\footnote{For more details about the TNG data products we refer  to \citet{2019ComAC...6....2N} and to the Illustris-TNG website: \url{http://www.tng-project.org/data/docs/specifications/}} simulations  provide  $2\times2500^3$  resolution elements in a volume of (300 Mpc)$^3$, at 100 redshift snapshots between $z$=0 and $z$=20. For each snapshot at redshift $z<1$, there are $\sim$ 2500 groups with $10^{13}$ $M_\odot \le M_{200} \le 10^{15}$ $M_\odot$, where $M_{200}$ and other parameters\footnote{As discussed by \cite{2017MNRAS.465.3291W,2018MNRAS.473.4077P}, the limited mass/spatial resolutions of the simulations may introduce numerical convergence issues, and as a consequence stellar masses may be underestimated. Comparing TNG300-1 with 
TNG100-1, where the resolution is higher but the volume is lower,  \citet{2018MNRAS.475..648P} showed that stellar masses for galaxies in groups and clusters may be underestimated in TNG300-1 by a factor $<$ 1.4.
Since this does not affect the results in our analysis, we decided to use the original, not rescaled stellar masses.}
from the simulations are defined in  Table~\ref{tab:ITNGdef}. The BCG in each group is identified as the  most massive subhalo.
To obtain a sample of simulated clusters matching as close as possible  those in the KiDS-\AMICO\ catalogue, clusters from the simulations are randomly selected so that they reproduce the distribution in  mass and redshift of real clusters.

\begin{table}
    \centering
	\caption{Definition of the group and subhalo fields from Illustris-TNG, used for the comparison with the observed quantities.}
	\label{tab:ITNGdef}
	\begin{tabular}{ccc}
		\hline
		Quantity & Name & Illustris-TNG\\
		\hline
		Cluster mass &	$M_{200}$& Group\_M\_Crit200 \\ 
		Cluster size &	$R_{200}$& Group\_R\_Crit200 \\ 
		Stellar mass  &	$M_{\rm star}$ & SubhaloStellarPhotometricsMassInRad \\
		\hline
	\end{tabular} 
\end{table}

We define the red and blue samples of galaxies in the simulation as we did for the real data,  that is according to the $g-r$ ($z\le0.4$) and  $r-i$ ($z>0.4$) colours produced by the {\em E} and {\em Sa} models. 
The comparison of the simulated and observed red and blue cluster members should be limited to few $R_{200}$ since the halo finder algorithm adopted in these simulations is designed to identify substructures on scales close to the virial radius \citep[see e.g.][]{2020A&A...639A.122K}.
Within $R_{200}$, density profiles (Figure~\ref{fig:mfrac_zbin}) and red fractions (Figure~\ref{fig:clz_mass_type})  show  an increasing disagreement for increasing redshift and mass values.  Consistently with what we find in our catalogue, the red fraction increases with increasing mass and decreasing redshift. However, while at low redshift (and low cluster masses) the agreement is good, at higher redshift ($z>0.4$) the simulated red fraction is much higher than what measured in our catalogue. For instance, in the redshift bin 0.6 $<$ z $<$ 0.8 the blue fraction is higher than the red fraction for all cluster masses in our catalogue, while this only happens for low mass groups in the simulations.

\begin{figure}
	\includegraphics[clip,width=1\linewidth]{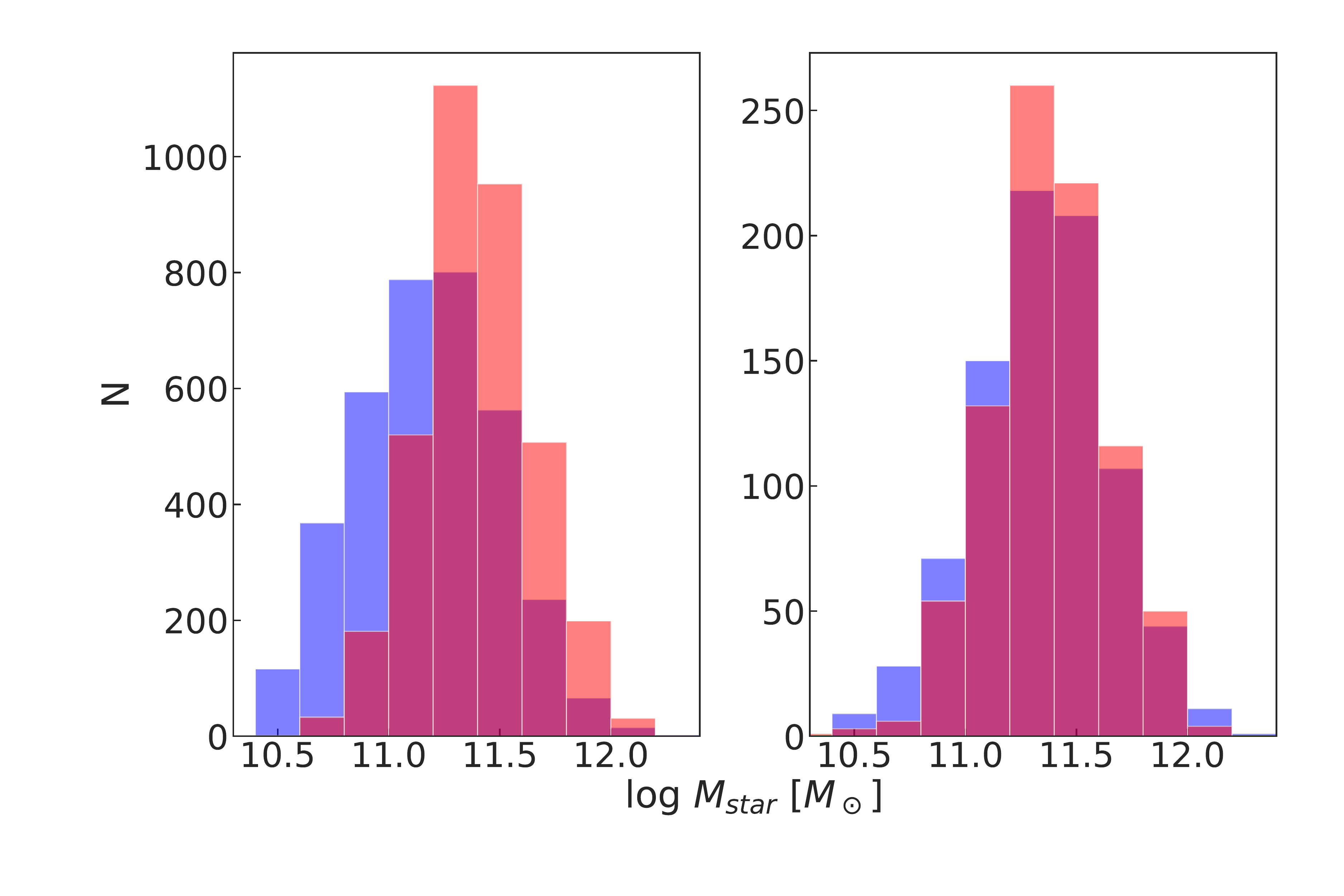}
	\caption{Distribution of BCG stellar masses in KiDS-\AMICO\   (blue),  the Illustris TNG300-1 sample (light red). In the {\em left} panel the Illustris TNG300-1 simulations are matched in cluster mass and redshift to the KiDS-\AMICO\  subsample with measured BCG stellar mass; in the {\em right} panel a further selection of clusters with $\Delta M_{1,2}>1$ is done.} 
	  \label{fig:stmass_A2tng}
\end{figure}

To compare simulated and measured stellar masses, we repeat the procedure described above, but taking as input the KiDS-\AMICO\  subsample for which a stellar mass measurement is available. The left panel of Figure~\ref{fig:stmass_A2tng} compares the distribution of simulated and measured BCG stellar masses, showing an excess of BCGs with measured $\log M_{\rm star}/M_\odot < 11$. 
An improved agreement is obtained further  selecting only clusters with $\Delta M_{1,2} >1$ (Figure~\ref{fig:stmass_A2tng}, right panel), confirming that clusters with a lower BCG stellar mass are also those with a lower magnitude gap.

\begin{figure}
	\includegraphics[clip,width=1.02\linewidth]{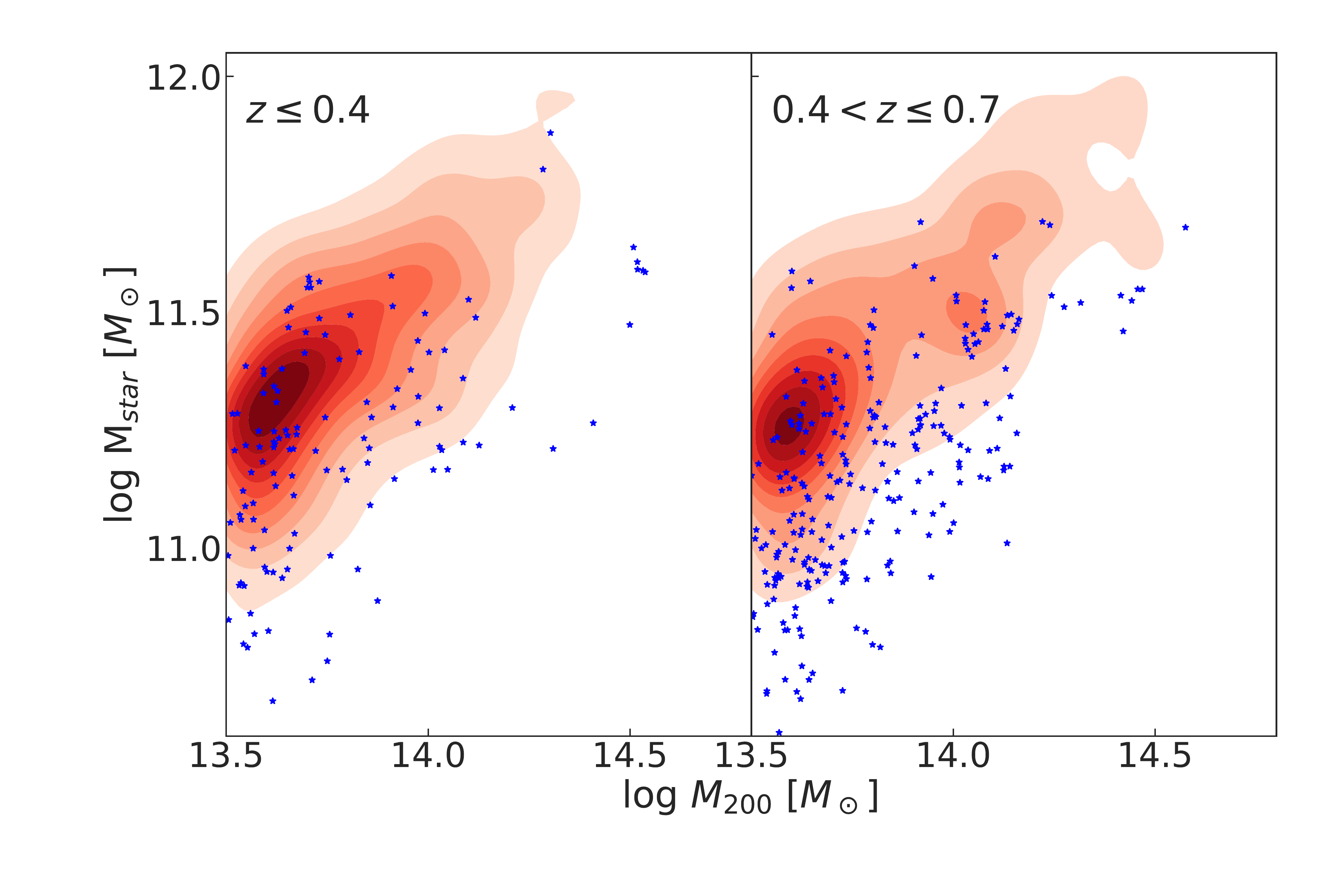}
	\caption{BCG stellar masses vs. cluster masses in the Illustris TNG300-1 sample: red contours display the distribution for red BCGs, values for blue BCGs are plotted as blue stars.
		\label{fig:Mstar2M200_itng}}
\end{figure}

Of the simulated BCGs, $\sim$ 10\%  are classified as blue, with a redshift distribution similar to what observed in our catalogue.
Based on \citet{2019MNRAS.487.5416T} who analyzed morphology, star formation and $g-r$ colours in TNG100 simulations, galaxies with such $g-r$ colours should be mainly disc, star-forming galaxies. Figure~\ref{fig:Mstar2M200_itng} shows the distribution of  cluster and BCG stellar masses for the  red/blue simulated clusters: they  are closer than real clusters (see Figure~\ref{fig:mbcg}) to the position occupied by relaxed clusters. Consistently with the observations, also in the simulations  the  stellar mass in blue BCGs is lower than in red BCGs at a given value of the cluster mass, though the stellar mass of observed blue BCGs is  $\sim 0.2$ dex lower than in simulated blue BCGs.

\section{Discussion and Conclusions}
\label{sec:conclusions}
In this paper we explored the properties (red/blue fraction of BCGs and member galaxies; cluster mass vs. the stellar mass, luminosity and size of the BCG) of galaxy clusters in the KiDS-\AMICO\  (DR3) and their evolution in redshift, based on a catalogue of $\sim$ 8000 clusters detected in the 414 deg$^2$ area covered by KiDS-DR3. Membership probabilities, which are used in our analysis, were validated by  the comparison with  spectroscopic redshifts available from the SDSS and the GAMA surveys. 
The comparison with clusters in the SDSS-{\em redMaPPer} catalogue selected in the same area and in the same cluster mass and redshift range shows that KiDS-\AMICO\ detects  more clusters with a lower red fraction than  {\em redMaPPer} and a cluster mass $\le 10^{14} M_\odot$.

The main results can be summarized as follows:
\begin{itemize}
 \item  At low redshifts ($z\le0.4$) clusters are dominated by red galaxies; the red fraction and trend with cluster mass and redshift are in good agreement with those obtained from the Illustris TNG300-1 simulations.  
\item At higher redshifts, the decrease with redshift in the cluster red fraction, implying an increase in the star formation,  is much faster in the real data than in the simulations. 
\item Another disagreement is seen in the cluster mass vs. BCG stellar mass distribution: at all redshifts the simulated data are closer to the position occupied in real data by relaxed clusters, while many clusters in the KiDS-\AMICO\ catalogue  show evidence for a dynamically disturbed status (low stellar mass at a fixed cluster mass and $\Delta M_{1,2}<1$).
\item  In blue  BCGs the  stellar mass is lower than in red BCGs for clusters of the same mass: this is also seen in the simulations, though  the difference is not as high as in the real data.
\end{itemize}

The difference in stellar mass between blue (star forming) and red (quiescent) BCGs probably reflects a different contribution by quenching in different cluster environments, in particular by ram pressure stripping  \citep{2019MNRAS.488.5370L},
with tidal events \citep{2020A&A...638A.133L} that  may trigger star formation but also deplete the stellar mass.

Based on multi-wavelength observations of clusters selected in the South-Pole Telescope Survey,  \citet{2016ApJ...817...86M} found that the  fraction of Star-Forming BCGs is rapidly increasing with redshift, with 20\% at  $z\sim0.4$ and $\sim$90\% at $z\sim$1 showing strong star formation (> 10 $M_\odot$ yr$^{-1}$), and that at $z>0.6$  they are found in morphologically disturbed clusters. This would be consistent with what we observe in our data, that is blue BCGs are found preferentially  in clusters at higher redshifts ($z>0.4$) and with a low value of the magnitude gap, $\Delta M_{1,2} <1$, indicative of a disturbed cluster dynamical status. There is evidence \citep{2019MNRAS.483.4140R} that  the Illustris-TNG simulations may not be yet able to fully reproduce gas-rich mergers: this may explain why simulated clusters show properties more typical of relaxed clusters, and as a 
consequence a lower fraction of star forming galaxies at increasing 
redshifts compared to observations.

Work is in progress to extend the cluster detection with \AMICO\ to the next KiDS data releases, that will give an increase in the survey area by a factor of $\sim$ 3.
At the same time, this will also increase the number of galaxies with measured structural parameters, allowing a more detailed analysis using this classification.

\section*{Acknowledgements}

We thank the anonymous referee for the  useful comments that improved the paper.

MS acknowledges financial contribution from contract ASI-INAF n.2017-14-H.0 and INAF `Call per interventi aggiuntivi a sostegno della ricerca di main stream di INAF'.  CT acknowledges funding from the INAF PRIN-SKA 2017 program 1.05.01.88.04.
LM acknowledges the support from the grant PRIN-MIUR 2017 WSCC32 and ASI n.2018-23-HH.0

Based on data products from observations made with ESO Telescopes
at the La Silla Paranal Observatory under programme IDs 177.A-3016, 177.A3017 and 177.A-3018, and on data products produced by Target/OmegaCEN, INAF-OACN, INAF-OAPD and the KiDS production team, on behalf of the KiDS consortium. GAMA is a joint European-Australasian project based around a spectroscopic campaign using the Anglo-Australian Telescope. The GAMA input catalogue is based on data taken from the Sloan Digital Sky Survey and the UKIRT Infrared Deep Sky Survey. Complementary imaging of the GAMA regions is being obtained by a number of independent survey programmes including GALEX MIS, VST KiDS, VISTA VIKING, WISE, Herschel-ATLAS, GMRT and ASKAP providing UV to radio coverage. GAMA is funded by the STFC (UK), the ARC (Australia), the AAO, and the participating institutions. The GAMA website is http://www.gama-survey.org/. 
We acknowledge the usage of the Illustris-TNG simulations and of the JupyterLab environment that was made available by the TNG Collaboration to explore and analyze the TNG simulations. 
We acknowledge the usage of the 
\texttt{StatsModels}, \texttt{AstroPy}, \texttt{bootstrapped}, \texttt{Pandas} and \texttt{Seaborn} libraries in Python.

\section*{Data availability}
The data underlying this article will be shared on reasonable request to the corresponding author.




\bibliographystyle{mnras}
\bibliography{paper} 








\bsp	
\label{lastpage}
\end{document}